\documentclass{jnmp03c}

\begin{document}
\setcounter{page}{106}
\newtheorem{lem}{Lemma}[section]
\newtheorem{theor}[lem]{Theorem}

\renewcommand{\evenhead}{S N M Ruijsenaars}
\renewcommand{\oddhead}{Reflectionless Analytic Difference Operators I.
Algebraic Framework}

\thispagestyle{empty}

\FistPageHead{1}{\pageref{ruijsenaars-firstpage}--\pageref{ruijsenaars-lastpage}
}{Article}

\copyrightnote{2001}{S N M Ruijsenaars}

\Name{Reflectionless Analytic Difference Operators \\ I.
Algebraic Framework}\label{ruijsenaars-firstpage}

\Author{S N M RUIJSENAARS}

\Adress{Centre for Mathematics and Computer Science\\
P.O. Box 94079, 1090 GB Amsterdam, The Netherlands}

\Date{Received April 17, 2000; Accepted June 30, 2000}

\begin{abstract}
\noindent
We introduce and study a class of analytic difference
operators admitting reflectionless eigenfunctions. Our
construction of the class is patterned after the Inverse
Scattering Transform for the reflectionless self-adjoint
Schr\"odinger and Jacobi operators corresponding to KdV
and Toda lattice solitons.
\end{abstract}

\renewcommand{\theequation}{\thesection.\arabic{equation}}

\setcounter{equation}{0}

\section{Introduction}

This paper and its two companion papers Refs.~\cite{r=0II} and
\cite{r=0III} (from now on referred to as Parts~II and~III)
originate from our previous work on reflectionless analytic
difference operators of relativistic Calogero--Moser type,
cf.~Refs.~\cite{glf2,hilb}. In our conference contribution
Ref.~\cite{NEEDS99} we already discussed the possible existence of
an extensive class of self-adjoint reflectionless analytic
difference operators containing those studied in Ref.~\cite{hilb}.
The present series of papers serves to enlarge the scenario
envisaged in Ref.~\cite{NEEDS99}, and in particular confirms some
conjectures made there.

In this first part we restrict attention to the algebraic
aspects of a huge class of analytic difference operators
that admit reflectionless eigenfunctions. In Part~II we
show that these operators and eigenfunctions
can be tied in with a non-local soliton evolution equation, and with
integrable (classical) $N$-body systems of relativistic
Calogero--Moser type.  Part~III is concerned with a
restricted class, for which we are able to prove
self-adjointness of the analytic difference operators by
exploiting various features of the reflectionless
eigenfunctions.

We proceed by delineating the analytic difference
operators at issue in this paper. They are of the form
\be\label{defA}
A=T_i+V_a(x)T_{-i}+V_b(x).
\ee
Here, the coefficients (``potentials'') $V_a(x)$ and
$V_b(x)$ are functions from the field ${\cal M}$ of
meromorphic functions, and the translations $T_{\pm i}$
are defined by
\be\label{defT}
(T_{\alpha}f)(x)\equiv f(x-\alpha),\qquad  \alpha \in
{\mathbb C}^{*},\qquad  f\in{\cal M}.
\ee
We only consider potentials satisfying
\be\label{Vas}
\lim_{|{\rm Re}\, x|\to \infty} V_a(x)=1,\qquad
 \lim_{|{\rm Re}\, x|\to \infty} V_b(x)=0.
\ee
Hence the analytic difference operator $A$ becomes equal
to the ``free'' analytic difference operator (henceforth
A$\Delta$O)
\be\label{A0}
A_0\equiv T_i+T_{-i}
\ee
for $|{\rm Re}\,  x|\to \infty$.

Obviously, $A_0$ admits plane wave eigenfunctions $\exp
(ixp)$ with eigenvalue $e^p+e^{-p}$ for all $p\in{\mathbb C}$. It
is therefore a natural question whether the A$\Delta$O $A$,
viewed as a linear operator on ${\cal M}$, admits
eigenfunctions with eigenvalue $e^p+e^{-p}$ and plane
wave asymptotics for $|{\rm Re}\,  x| \to \infty$. More
specifically, this question reads: Do there exist
functions ${\cal W} (\cdot ,p)\in {\cal M}$, satisfying
\be\label{Aeig}
A{\cal W} (x,p)=(e^p+e^{-p}){\cal W}(x,p),
\ee
\be\label{W+}
{\cal W}(x,p)\sim \exp (ixp),\qquad {\rm Re}\,  x\to\infty,
\ee
\be\label{W-}
{\cal W}(x,p)\sim a(p)\exp (ixp)+b(p)\exp (-ixp),\qquad {\rm Re}\,
x\to -\infty,
\ee
for generic $p\in {\mathbb C}$?

To our knowledge, this question has not been addressed in previous
literature. Here we do not answer it either. However, we do
present a vast class of potential pairs $V_a$, $V_b$ for which we
answer the question in the affirmative. Moreover, the pertinent
eigenfunctions are reflectionless, in the sense that in (\ref{W-})
one has $b(p)=0$. They are of the form \be\label{Wform} {\cal
W}(x,p)=\exp(ixp)\left( 1-\sum_{n=1}^N
\frac{R_n(x)}{e^p-z_n}\right), \ee where $z_1,\ldots,z_N$ are
distinct complex numbers, and $R_1,\ldots,R_N\in{\cal M}^{*}$.
(Here and below, ${\cal M}^{*}$~denotes the space of meromorphic
functions with the zero function deleted.)

It should be emphasized at the outset that whenever such
eigenfunctions exist, they are highly non-unique. Indeed,
introducing the infinite-dimensional space \be\label{cPa} {\cal
P}_{\alpha}\equiv \{ \mu \in{\cal M}^{*} \mid \mu (x+\alpha )=\mu
(x)\} , \qquad \alpha\in{\mathbb C}^{*}, \ee of $\alpha$-periodic
multipliers, one  verifies first of all that whenever ${\cal
W}(x,p_0)$ satisfies (\ref{Aeig}) for a certain $p_0\in{\mathbb
C}$, then the function $\mu(x){\cal W}(x,p_0)$ also satisfies
(\ref{Aeig}) for all $\mu(x)\in{\cal P}_i$. Moreover, introducing
\be\label{cPac} {\cal P}_{\alpha}(c)\equiv \left\{ \mu\in{\cal
P}_{\alpha}\Bigl| \lim_{|{\rm Re}\, x|\to \infty} \mu(x)=c,\
c\in{\mathbb C}\right\}, \ee and choosing $\mu\in{\cal P}_i(1)$,
the function $\mu(x){\cal W}(x,p_0)$ has once more plane wave
asymptotics for $|{\rm Re}\,  x|\to\infty$. Thus, when we
supplement the numbers $z_1,\ldots,z_N$ occurring in (\ref{Wform})
with further complex numbers $z_{N+1},\ldots,z_M$, such that
$z_1,\ldots,z_M$ are distinct, then the function \be\label{Wamb}
\tilde{{\cal W}}(x,p)\equiv{\cal W} (x,p) \left( 1-\sum_{n=N+1}^M
\frac{\nu_n(x)}{e^p-z_n}\right), \qquad  \nu_n\in{\cal P}_i(0),
\ee is of the same form, and satisfies (\ref{Aeig})--(\ref{W-}) as
well.

For the class of potentials involved, however, this non-uniqueness
can be obviated by restricting attention to ``residue functions''
$R_1(x),\ldots,R_N(x)$ with quite special pro\-per\-ties. (This is
shown in Lemma~4.2 below.) In point of fact, we will first define
functions $R_1,\ldots,R_N\in{\cal M}^{*}$ as the unique solutions
to a certain system of $N$ linear equations, given by
(\ref{sysN}). This system involves suitably constrained complex
numbers $z_1=\exp(r_1),\ldots,z_N=\exp(r_N)$, cf.~(\ref{req1}),
(\ref{req2}), and multipliers $\mu_1(x),\ldots,\mu_N(x)$
sa\-tisfying (\ref{req3}). Subsequently, we define a function
${\cal W}(x,p)$ by (\ref{Wform}). Then we show that this function
has asymptotics (\ref{W+}) and (\ref{W-}) with $b(p)=0$, and prove
that there exists a uniquely determined A$\Delta$O $A$ of the
above type satisfying (\ref{Aeig}) for $\exp(p)\ne
z_1,\ldots,z_N$. (For this discrete set of $p$-values the wave
function ${\cal W}(x,p)$ (\ref{Wform}) is of course ill defined.)

The procedure just sketched will be detailed in Section~2. It owes
much to the account of reflectionless Schr\"odinger operators that
can be found in Newell's monograph Ref.~\cite{newe}. More
generally, it is inspired by the inverse scattering transform
(IST) for one-dimensional self-adjoint Schr\"odinger and Jacobi
operators. As is well known (see, for instance,
Refs.~\cite{newe}--\cite{fata}), in the reflectionless case one
winds up with a linear system of $N$ equations, where~$N$ is the
number of bound states. This system involves the Cauchy matrix
$C$, just as the system (\ref{sysN}). (We have relegated the
features of $C$ we need to Appendix~A.)

Our class of A$\Delta$Os admitting reflectionless
eigenfunctions is far more extensive than the
reflectionless Schr\"odinger and Jacobi operators
obtained via the IST. This is because we can allow
arbitrary multiplier functions $\mu_n(x)$ from the
infinite-dimensional space ${\cal P}_i(c_n),n=1\ldots,N$,
cf.~(\ref{req3}). Choosing $\mu_n(x)$ equal to the
constant $c_n$ for all $n\in\{ 1,\ldots ,N\}$, it is
still ``twice as large'', in the sense that we neither
require $|z_n|=1$ nor conditions on $c_n$.

On the other hand, we do require constant multipliers and
reality conditions in Part~III of this series, which
deals with Hilbert space aspects~\cite{r=0III}. Indeed,
we need such constraints in order to exploit the wave
function ${\cal W}(x,p)$ for associating to the A$\Delta$O $A$ a
bona fide self-adjoint operator $\hat{A}$ on the Hilbert
space $L^2({\mathbb R},dx)$. (To date, a general self-adjointness
theory in which our indirect definition of $\hat{A}$ fits
is not available.)

For our present purposes it suffices to mention one
prominent fact illustrating the considerable differences
between reflectionless self-adjoint A$\Delta$Os and
reflectionless self-adjoint Schr\"odinger and Jacobi
operators. This is the existence of an
infinite-dimensional family of self-adjoint
reflectionless A$\Delta$Os without {\em any} bound states.
(Actually, a smaller, but still infinite-dimensional
family of such A$\Delta$Os can already be found in our
paper Ref.~\cite{hilb}, cf.~also Ref.~\cite{NEEDS99}.) This
circumstance shows that one should not expect a
straightforward analog of the direct transform associated
with the Schr\"odinger and Jacobi cases, and indeed we
have very little to say about the direct problem for
A$\Delta$Os.

We continue with a more detailed account of the
organization and results of this paper. As
mentioned above, Section~2 is concerned with the groundwork
for this series of papers. Fixing $N\in{\mathbb N}^{*}$, the
multipliers (\ref{req3}) already form an
infinite-dimensional family, yet they are by no means the
largest class giving rise to A$\Delta$Os admitting
reflectionless eigenfunctions. To illustrate the freedom
involved, we have made a close-up of the $N=1$ case, where
all pertinent objects can be inspected without difficulty.

The restrictions (\ref{req1}) and (\ref{req2}) are primarily
motivated by the applications in Parts~II and III, where they will
be seen to be quite natural. On the other hand, for the soliton
solutions in Part~II and for all of Part~III we also need the
multipliers to be constant. It is a striking feature of the
algebraic viewpoint adopted in this first part that the freedom
allowed by (\ref{req3}) does not give rise to any additional
difficulty, as compared to allowing only constants. (By contrast,
when we try to relax the requirements (\ref{req1}), (\ref{req2})
and~(\ref{req3}), we typically find that some of our arguments
break down.) In particular, one obtains the same transmission
coefficient (given by (\ref{af})) for all of these multipliers.

Section~3 contains on the one hand some general insights
bearing on the question whether the potential $V_b(x)$
can vanish identically. On the other hand, these results
point the way towards special cases where one does have
$V_b=0$. This enables us to show that when all of the
numbers $r_1,\ldots,r_N$ in (\ref{req1}) have imaginary
parts in $(0,\pi)$ or in $(-\pi,0)$, then the wave
function is not only an eigenfunction of $A$
(\ref{defA}) with eigenvalue $\exp(p)+\exp(-p)$, but
also an eigenfunction with eigenvalue
$\exp(p/2)+\delta\exp(-p/2)$ of an A$\Delta$O
\be
S_{\delta}=T_{i/2}+\delta V(x)T_{-i/2},\qquad \delta =+,-.
\ee
Here, one has
\be
\lim_{|{\rm Re}\, x|\to \infty} V(x)=1,
\ee
and $\delta =+/-$ corresponds  to ${\rm Im}\, r_1,\ldots,{\rm Im}\, r_N$
belonging to $(0,\pi)$/$(-\pi,0)$, resp.

Since we {\em construct} reflectionless wave functions
${\cal W}(x,p)$ and A$\Delta$Os $A$ related via (\ref{Aeig}) from
given data $(r,\mu)$ satisfying (\ref{req1}),
(\ref{req2}) and (\ref{req3}), it is an obvious question
whether distinct data $(r,\mu)$ can give rise to the same
${\cal W}(x,p)$ and/or $A$. Section~4 is devoted to a study of
this injectivity problem. As it turns out, it is rather
easy to answer the question completely for ${\cal W}(x,p)$,
cf.~Theorem~4.1. For $A$, however, our results are not
complete. But the partial answers we obtain in
Lemmas~4.2--4.4 yield considerable evidence for our
conjecture that the ``generalized IST map'' from $(r,\mu)$
to $A$ is injective up to permutations.

This paper is concluded with four Appendixes. In
Appendix~A we collect some information on matrices of
Cauchy and Vandermonde type that occur in the main text.
Appendix~B concerns the Casorati determinants associated
to the solutions of the ordinary second-order analytic
difference equations at hand. Its results are of interest
in itself. They are also a crucial input for our study of
the injectivity problem in Section~4.

In Appendix~C we obtain alternative representations for various
important quantities, including the wave function and potentials.
In Appendix~D we use these formulas to show that the A$\Delta$O
$A$ (\ref{defA}) is {\em formally} self-adjoint on $L^2({\mathbb
R},dx)$, provided the numbers $r_n$, $n=1,\ldots,N$, are purely
imaginary and the functions $i\exp(-r_n)\mu_n(x)$, $n=1,\ldots,N$,
are real-valued for real $x$.


\setcounter{equation}{0}

\section{Reflectionless A$\pbf{\Delta}$O-eigenfunctions}

Consider a function of the form (\ref{Wform}), with
$R_1,\ldots,R_N\in{\cal M}^{*}$ and $z_1,\ldots,z_N$ distinct
numbers in ${\mathbb C}^{*}$, written as \be z_n\equiv \exp
(-r_n),\qquad {\rm Im}\, r_n \in (-\pi,\pi],\qquad n=1,\ldots,N,
\ee but without further constraints for the moment. Introducing
the auxiliary wave function \be\label{defaux} {\cal
A}(x,p)=\prod_{n=1}^N\left(e^p-e^{-r_n}\right)\cdot {\cal W}(x,p),
\ee it follows from (\ref{Wform}) that ${\cal A}(x,p)$ can be
rewritten as \be\label{auxform} {\cal A}(x,p)=e^{ixp}
\left(e^{Np}+\sum_{k=0}^{N-1}c_k(x)e^{kp}\right),\qquad
c_k\in{\cal M}. \ee Here we have in particular
\be\label{ca}c_0(x)=\prod_{n=1}^N\left(-e^{-r_n}\right)\cdot\lambda(x),\qquad
 \lambda(x)\equiv 1+\sum_{n=1}^Ne^{r_n}R_n(x),
\ee
\be\label{cb}
c_{N-1}(x)=\sum_{n=1}^N\left(-e^{-r_n}-R_n(x)\right).
\ee

Let us now ask: When is ${\cal A}(x,p)$ an eigenfunction with
eigenvalue $e^p+e^{-p}$ of an A$\Delta$O~$A$ of the form
(\ref{defA})? Clearly, this amounts to ${\cal A}(x,p)$ satisfying
the analytic difference equation (from now on A$\Delta$E)
\be\label{Aade} {\cal A}(x-i,p)+V_a(x){\cal
A}(x+i,p)+\left(V_b(x)-e^p-e^{-p}\right){\cal A}(x,p)=0, \ee for
certain functions $V_a,V_b\in{\cal M}$. In view of the simple
structure (\ref{auxform}) of ${\cal A}(x,p)$, this comes down to a
system of $N+2$ equations relating the functions
$c_0,\ldots,c_{N-1}$ and potentials $V_a$, $V_b$. In particular,
the vanishing of the coefficients of $e^{Np}$ and $e^{-p}$ is
equivalent to the relations \be\label{Vbc}
c_{N-1}(x-i)+V_b(x)-c_{N-1}(x)=0, \ee \be\label{Vac}
V_a(x)c_0(x+i)-c_0(x)=0. \ee Therefore, $V_b(x)$ is uniquely
determined by ${\cal A}(x,p)$, and so is $V_a(x)$, provided
$\lambda(x)$ does not vanish identically.

Of course, for $N=0$ this reasoning yields the unsurprising
consequence $V_a(x)=1$, $V_b(x)=0$. But already for $N=1$ one gets
in addition to (\ref{Vbc}) and (\ref{Vac}) the relation
\be\label{N1a} V_a(x)+V_b(x)c_0(x)-1=0\qquad  (N=1). \ee Using
(\ref{Vbc}) and (\ref{Vac}), this amounts to $c_0(x)$ satisfying
\be\label{N1b}
\frac{c_0(x)}{c_0(x+i)}+[c_0(x)-c_0(x-i)]c_0(x)=1
\qquad (N=1). \ee
Thus, even in this simple case the extra constraint looks
forbidding.

\strut\hfill

More generally, when one starts from a function ${\cal A}(x,p)$ of
the form (\ref{auxform}) with $c_0(x)\in{\cal M}^{*}$, and defines
$V_a$ and $V_b$ by \be\label{Vadef} V_a(x)\equiv c_0(x)/c_0(x+i),
\ee \be\label{Vbdef} V_b(x)\equiv c_{N-1}(x)-c_{N-1}(x-i), \ee it
seems hopeless to solve the $N$ remaining nonlinear A$\Delta$Es
for the coefficients $c_0,\ldots,c_{N-1}$.

Even so, we are going to construct a large class of solutions by
viewing ${\cal A}(x,p)$ as arising from (\ref{Wform}) via
(\ref{defaux}). We have already seen that these formulas determine
the coefficients~$c_k$ in (\ref{auxform}) in terms of complex
numbers $\exp(-r_1),\ldots,\exp(-r_N)$ and ``residue functions''
$R_1,\ldots,R_N$. As it turns out, $R_1,\ldots,R_N$ can be defined
via a linear system of~$N$ equations such that (\ref{Aade}) is
obeyed, provided $V_a$ is defined by (\ref{Vadef}) and (\ref{ca}),
and~$V_b$ by (\ref{Vbdef}) and (\ref{cb}).

It is illuminating to present the details first for $N=1$. Using
(\ref{cb}) (or (\ref{ca})) to express~$c_0$ in terms of the
residue function $R\equiv R_1$ at the pole $p=-r\equiv -r_1$ of
${\cal W}(x,p)$, we begin by noting \be\label{A-r} {\cal
A}(x,-r)=-e^{-irx}R(x), \ee \be\label{A+r} {\cal
A}(x,r)=e^{irx}\left(e^r-e^{-r}-R(x)\right). \ee For the pertinent
linear constraint on $R$ we now need to require $e^{2r}\ne 1$, or,
equivalently, $r\ne 0,i\pi$. Then it reads \be\label{Rcon}
\mu(x)e^{-2irx}R(x)+\frac{1}{e^r-e^{-r}}R(x)=1, \ee where $\mu(x)$
is an arbitrary function in ${\cal P}_i$ (\ref{cPa}). Obviously,
(\ref{Rcon}) determines a unique function $R\in{\cal M}^{*}$.

The crux of the constraint (\ref{Rcon}) is that it guarantees
that ${\cal A}(x,p)$ fulfils
\be\label{Arel}
{\cal A}(x,r)=\left(e^{-r}-e^{r}\right)\mu(x){\cal A}(x,-r),\qquad
\mu\in{\cal P}_i,
\ee
as is clear from (\ref{A-r}), (\ref{A+r}). Next, we observe
that a function ${\cal E}(x,p)$ of the form
\be\label{E1}
{\cal E}(x,p)=e^{ixp}\left(e^p+c(x)\right),
\ee
is uniquely determined when it satisfies
\be\label{E2}
{\cal E}(x,r)=\left(e^{-r}-e^r\right)\mu(x){\cal E}(x,-r),\qquad
\mu\in{\cal P}_i.
\ee
Indeed, substituting (\ref{E1}) in
(\ref{E2}), one gets a linear constraint that
uniquely determines $c\in{\cal M}^{*}$.

We are now going to exploit the uniqueness of ${\cal E}$. With
$R(x)$ determined by (\ref{Rcon}), we define $V_a$ and $V_b$ via
(\ref{Vadef}), (\ref{Vbdef}) and (\ref{cb}). Consider the function
${\cal D}(x,p)$ on the lhs of~(\ref{Aade}). By construction, it is
of the form \be {\cal D}(x,p)=e^{ixp}d(x),\qquad d\in{\cal M}. \ee
Moreover, on account of (\ref{Arel}) and the $i$-periodicity of
$\mu(x)$, it obeys \be {\cal
D}(x,r)=\left(e^{-r}-e^r\right)\mu(x){\cal D}(x,-r). \ee But then
the function ${\cal E} ={\cal A} -{\cal D}$ has the two features
(\ref{E1}), (\ref{E2}) that uniquely determine~${\cal A}$.
Therefore, $d(x)$ must vanish identically, so that ${\cal A}(x,p)$
solves the A$\Delta$E (\ref{Aade}). (In particular, this entails
that $c_0(x)=-\exp(-r)-R(x)$ solves the nonlinear A$\Delta$E
(\ref{N1b}).)

The upshot is that the wave function
\be
{\cal W}(x,p)=e^{ixp}\left( 1-\frac{R(x)}{e^p-e^{-r}}\right),
\ee
with $\exp(2r)\ne 1$ and $R\in{\cal M}^{*}$ given by (\ref{Rcon}),
satisfies the eigenvalue equation (\ref{Aeig}), with $V_a$ and
$V_b$ given by
\be
V_a(x)=\frac{e^{-r}+R(x)}{e^{-r}+R(x+i)},
\ee
\be
V_b(x)=R(x-i)-R(x).
\ee
But we need further constraints to guarantee the asymptotics
(\ref{Vas}), (\ref{W+}) and (\ref{W-}).

Clearly, (\ref{W+}) and (\ref{W-}) are equivalent to
\be\label{Ras+}
\lim_{{\rm Re}\, x\to \infty} R(x)=0,
\ee
\be\label{Ras-}
\lim_{{\rm Re}\, x\to -\infty} R(x)=\kappa,\qquad \kappa\in{\mathbb C}.
\ee
Moreover, (\ref{Ras+}) and (\ref{Ras-}) imply (\ref{Vas}),
provided $\kappa\ne -\exp(-r)$. But (\ref{Vas}) can be fulfilled
without (\ref{Ras+}) holding true. For example, choosing
$\mu(x)$ equal to $1/\cosh (4\pi x)$, one gets $R(x)\to
\exp(r)-\exp(-r)\ne 0$ as $|{\rm Re}\,  x|\to \infty$,
cf.~(\ref{Rcon}).

In the example just given the function $\mu(x)\exp(-2irx)$ has
limit 0 for $|{\rm Re}\,  x|\to \infty$. But we can also let it
diverge for $|{\rm Re}\,  x|\to\infty$, by choosing for instance
$\mu(x)=\cosh (4\pi x)$. Then we obtain $R(x)\to 0$ as $|{\rm Re}\,
x|\to\infty$, so that (\ref{Vas}), (\ref{W+}) and (\ref{W-})
are satisfied, with $a(p)=1$ and $b(p)=0$. As a result, already
for
$N=1$ we obtain a huge class of A$\Delta$Os of the form
(\ref{defA})--(\ref{Vas}) admitting eigenfunctions ${\cal W}(x,p)$
of the form (\ref{Wform}) with {\em trivial} scattering.

Consider next the case
\be\label{c+}
\mu(x)\in{\cal P}_i(c),\qquad c\ne 0,\qquad {\rm Im}\, r\in(0,\pi).
\ee
Then we read off from (\ref{Rcon}) that
\be\label{Rdif}
\lim_{{\rm Re}\, x\to \infty} R(x)=0,\qquad
 \lim_{{\rm Re}\, x\to -\infty} R(x)=e^r-e^{-r}.
\ee
Therefore, (\ref{Vas}), (\ref{W+}) and (\ref{W-}) are once
again satisfied, now with $b(p)=0$ and
\be\label{aN1}
a(p)=\frac{e^p-e^r}{e^p-e^{-r}}.
\ee

There is yet another way to obtain (\ref{Rdif}) and hence
(\ref{Vas}), (\ref{W+}), (\ref{W-}), with $b(p)=0$ and $a(p)$
(\ref{aN1}). Indeed, we may choose \be\label{c-} e^{-2\pi x}
\mu(x)\in{\cal P}_i(c),\qquad c\ne 0,\qquad {\rm Im}\,
r\in(-\pi,0). \ee Observe that for real $p$ the functions $a(p)$
thus obtained are the complex conjugates of the functions $a(p)$
obtained from the choice (\ref{c+}).

We are singling out the cases (\ref{c+}) and (\ref{c-}), since our
requirements for {\em arbitrary}~$N$ reduce to (\ref{c+}) and
(\ref{c-}) for $N=1$. Indeed, turning to the general~$N$ case, our
conditions on the numbers $r_1,\ldots,r_N$ read as follows. First,
one has \be\label{req1} {\rm Im}\, r_n\in (-\pi,0)\cup
(0,\pi),\qquad n=1,\ldots, N. \ee Thus we have $N_{+}$ complex
numbers in the strip $\{ {\rm Im}\, r\in (0,\pi)\}$ and $N_{-}$ in
$\{ {\rm Im}\,  r\in (-\pi,0)\}$, with \be\label{NN} N_{+}\in \{
0,1,\ldots,N\},\qquad N_{-}=N-N_{+}. \ee Second, we require
\be\label{req2} e^{r_m}\ne e^{\pm r_n},\qquad 1\le m<n\le N. \ee
The conditions (\ref{req1}) and (\ref{req2}) ensure that the
Cauchy matrix \be\label{Cr} C(r)_{mn}\equiv
\frac{1}{e^{r_m}-e^{-r_n}},\qquad m,n=1,\ldots,N, \ee is well
defined and regular, cf.~Appendix~A.

Next, we choose multipliers $\mu_1(x),\ldots,\mu_N(x)$,
satisfying
\be\label{req3}
\mu_n\in{\cal P}_i(c_n),\qquad c_n\in{\mathbb C}^{*},\qquad n=1,\ldots,N.
\ee
Then we introduce a diagonal matrix
\be\label{defD}
D(r,\mu;x)\equiv {\rm diag}\,  (d(r_1,\mu_1;x),\ldots,
d(r_N,\mu_N;x)),
\ee
where the function $d$ is defined by
\be\label{defd}
d(\rho,\nu;x)\equiv \left\{
\begin{array}{ll}
\nu(x)e^{-2i\rho x},  &  {\rm Im}\,  \rho \in(0,\pi),  \vspace{1mm}\\
\nu(x)e^{-2i(\rho +i\pi)x},  & {\rm Im}\,  \rho\in(-\pi,0),
\end{array} \right. \qquad
\nu\in{\cal P}_i(c),\qquad c\in{\mathbb C}^{*}.
\ee

We are now prepared to study the linear system of $N$ equations
\be\label{sysN} (D(r,\mu;x)+C(r))R(x)=\zeta,\qquad \zeta\equiv
(1,\ldots,1)^t, \ee for $N$ unknown functions $R_1,\ldots,R_N$.
Let us begin by noting that the determinant \mbox{$|D(x)+C|$}
cannot vanish identically. Indeed, we have $|C|\ne 0$ and it is
clear from (\ref{defd}) that $D(x)\to 0$ for ${\rm Re}\,  x\to
-\infty$. (Here we suppressed the $(r,\mu)$-dependence, as we
often do in the sequel.)

Next, we denote by $Z_n(x)$ the matrix obtained upon replacing
the $n$th column of $D(x)+C$ by $\zeta$. Then we deduce that
the system (\ref{sysN}) admits a unique solution, given by
Cramer's rule:
\be\label{Rnform}
R_n(r,\mu;x)=|Z_n(r,\mu;x)|/|D(r,\mu;x)+C(r)|,\qquad
n=1,\ldots,N.
\ee
In the following lemma we collect some salient features of
$R(x)$.

\begin{lem}
The solution $R(r,\mu;x)$ (\ref{Rnform}) to the system
(\ref{sysN}) belongs to ${\cal M}^{*N}$. It satisfies
\be\label{R+}
\lim_{{\rm Re}\, x\to \infty} R(r,\mu ;x)=0,
\ee
\be\label{R-}
\lim_{{\rm Re}\, x\to -\infty} R(r,\mu ;x)=C(r)^{-1}\zeta,
\ee
and
\be\label{R+1}
\lim_{{\rm Re}\, x\to \infty} c_ne^{-2ir_nx}R_n(r,\mu ;x)=1,\qquad
  {\rm Im}\,  r_n\in(0,\pi),
\ee
\be\label{R+2}
\lim_{{\rm Re}\, x\to \infty} c_ne^{-2i(r_n+i\pi)x}R_n(r,\mu ;x)=1,\qquad
{\rm Im}\,
r_n\in(-\pi,0).
\ee
\end{lem}

\medskip

\noindent
{\bf Proof.} Obviously, (\ref{defd}) yields
\be\label{das-}
\lim_{{\rm Re}\, x\to -\infty} d(\rho,\nu ;x)=0,
\ee
\be\label{das+1}
\lim_{{\rm Re}\, x\to \infty} c^{-1}e^{2i\rho x}d(\rho ,\nu ;x)=1,  \qquad
{\rm Im}\,
\rho \in(0,\pi),
\ee
\be\label{das+2}
\lim_{{\rm Re}\, x\to \infty} c^{-1}e^{2i(\rho +i\pi)x}d(\rho ,\nu ;x)=1,
\qquad {\rm Im}\,
\rho \in(-\pi,0).
\ee
From (\ref{das-}) we deduce (\ref{R-}). In view of the
identities (\ref{Crid}), we have $\left(C(r)^{-1}\zeta\right)_n\ne 0$, so
that $R_n\in{\cal M}^{*}$. To prove (\ref{R+1}), (\ref{R+2}), we
rewrite (\ref{Rnform}) as
\be
R_n(x)=\left|Z_n(x)D(x)^{-1}\right|/\left|{\bf 1}_N+CD(x)^{-1}\right|,\qquad
n=1,\ldots,N.
\ee
When ${\rm Im}\,  r_n\in(0,\pi)$, we now multiply the $n$th column of
$Z_n(x)D(x)^{-1}$ by $c_n\exp(-2ir_nx)$ and use (\ref{das+1})
to obtain (\ref{R+1}). Likewise, (\ref{R+2}) follows from
(\ref{das+2}). Finally, (\ref{R+}) is evident from (\ref{R+1})
and (\ref{R+2}). \hfill \rule{3mm}{3mm}

\medskip

Next, we define the functions
\be\label{laf}
\lambda(r,\mu;x)\equiv 1+\sum_{n=1}^Ne^{r_n}R_n(r,\mu;x),
\ee
\be\label{af}
a(r;p)\equiv \prod_{n=1}^N\frac{e^p-e^{r_n}}{e^p-e^{-r_n}}.
\ee
Their pertinent properties are once more collected in a lemma.

\begin{lem}
The function $a(p)$ (\ref{af}) can be rewritten as
\be\label{aaux}
1-\sum_{n=1}^N\frac{\left(C(r)^{-1}\zeta\right)_n}{e^p-e^{-r_n}}.
\ee
The function $\lambda(x)$ (\ref{laf}) satisfies
\be\label{la+}
\lim_{{\rm Re}\, x\to \infty} \lambda(r,\mu;x)=1,
\ee
\be\label{la-}
\lim_{{\rm Re}\, x\to -\infty}
\lambda(r,\mu;x)=\exp\left(2\sum_{n=1}^Nr_n\right).
\ee
\end{lem}

\noindent
{\bf Proof.} The functions (\ref{aaux}) and (\ref{af}) are
$2i\pi$-periodic meromorphic functions of $p$ with finite
limits for ${\rm Re}\,  p\to -\infty$ and limit 1 for ${\rm Re}\,
p\to\infty$. Therefore, one need only verify equality of
residues at the simple poles $p=-r_m$, $m=1,\ldots,N$, to
conclude (by Liouville's theorem) that they coincide. This can
be done via the identities (\ref{Crid}).

Next, we note that (\ref{R+}) entails (\ref{la+}), whereas
(\ref{R-}) yields
\be
\lim_{{\rm Re}\, x\to -\infty}
\lambda(r,\mu;x)=1+\sum_{n=1}^Ne^{r_n}\left(C(r)^{-1}\zeta
\right)_n.
\ee
Taking ${\rm Re}\,  p\to -\infty$ in (\ref{aaux}) and (\ref{af}), we
obtain
\be
1+\sum_{n=1}^Ne^{r_n}\left(C(r)^{-1}\zeta\right)_n=\prod_{n=1}^Ne^{2r_n}.
\ee
Hence (\ref{la-}) results.\hfill \rule{3mm}{3mm}

\medskip

We now use these lemmas to study the potentials and A$\Delta$O
\be\label{Vaf}
V_a(r,\mu;x)\equiv\lambda(r,\mu;x)/\lambda(r,\mu;x+i),
\ee
\be\label{Vbf}
V_b(r,\mu;x)\equiv \sum_{n=1}^N (R_n(r,\mu;x-i)-R_n(r,\mu;x)),
\ee
\be\label{Af}
A(r,\mu)\equiv T_i+V_a(r,\mu;x)T_{-i}+V_b(r,\mu;x),
\ee
in relation to the wave function
\be\label{Wf}
{\cal W}(r,\mu;x,p)\equiv e^{ixp}\left( 1-\sum_{n=1}^N
\frac{R_n(r,\mu;x)}{e^p-e^{-r_n}}\right).
\ee

\begin{theor}
One has the limits
\be\label{Vasf}
\lim_{|{\rm Re}\, x|\to \infty} V_a(r,\mu;x)=1,\qquad
 \lim_{|{\rm Re}\, x|\to \infty} V_b(r,\mu;x)=0,
\ee
\be\label{W+f}
\lim_{{\rm Re}\, x\to \infty} e^{-ixp}{\cal W}(r,\mu;x,p)=1,
\ee
\be\label{W-f}
\lim_{{\rm Re}\, x\to -\infty} e^{-ixp}{\cal W}(r,\mu;x,p)=a(r;p),
\ee
where $a(r;p)$ is given by (\ref{af}). Furthermore, the wave
function (\ref{Wf}) satisfies the eigenvalue equation
\be\label{AWf}
A(r,\mu){\cal W}(r,\mu;x,p)=\left(e^p+e^{-p}\right){\cal W}(r,\mu;x,p),
\ee
for all $p$ with $\exp(p)\ne \exp(-r_1),\ldots,\exp(-r_N)$.
\end{theor}

\noindent
{\bf Proof.} From (\ref{R+}) and (\ref{la+}) it is obvious
that (\ref{Vasf}) holds for ${\rm Re}\,  x\to\infty$. Likewise,
(\ref{R-}) yields $V_b\to 0$ for ${\rm Re}\,  x\to -\infty$. The rhs
of (\ref{la-}) is non-zero, entailing $V_a\to 1$ for ${\rm Re}\,  x\to
-\infty$. Thus (\ref{Vasf}) is proved.

The limit (\ref{W+f}) is immediate from (\ref{R+}). From
(\ref{R-}) we obtain
\be
\lim_{{\rm Re}\, x\to -\infty}
e^{-ixp}{\cal W}(r,\mu;x,p)=1-\sum_{n=1}^N
\frac{\left(C(r)^{-1}\zeta\right)_n}{e^p-e^{-r_n}},
\ee
which equals $a(r;p)$, by virtue of Lemma~2.2. Thus it remains
to prove the eigenvalue equation (\ref{AWf}).

Clearly, (\ref{AWf}) will follow once we show that the
auxiliary wave function ${\cal A}(x,p)$ (\ref{defaux}) satisfies
the A$\Delta$E (\ref{Aade}). To this end we calculate
\be
{\cal A}(x,-r_m)=-e^{-ir_mx}\prod_{n\ne
m}\left(e^{-r_m}-e^{-r_n}\right)\cdot R_m(x),
\ee
\be
{\cal A}(x,r_m)=e^{ir_mx}\prod_{k=1}^N\left(e^{r_m}-e^{-r_k}\right)\cdot
 \left( 1-\sum_{n=1}^N
\frac{R_n(x)}{e^{r_m}-e^{-r_n}}\right), \ee and use (\ref{sysN})
to deduce \be\label{ArelN} {\cal
A}(x,r_m)=\left(e^{-r_m}-e^{r_m}\right) \!\prod_{n=1,\, n\ne
m}^N\!
\frac{\left(e^{r_m}-e^{-r_n}\right)}{\left(e^{-r_m}-e^{-r_n}\right)}
\cdot e^{2ir_mx}d(r_m,\mu_m;x){\cal A}(x,-r_m).\hspace{-5mm} \ee

Recalling the definition (\ref{defd}) of $d$, we see
that ${\cal A}(x,p)$ satisfies $N$ equations of the form
\be\label{keyeq}
{\cal A}(x,r_m)=\alpha_m(x){\cal A}(x,-r_m),\qquad m=1,\ldots,N,
\ee
\be\label{alm}
{\rm Im}\,  r_m \in \left\{
\begin{array}{l}(0,\pi) \vspace{1mm}\\
(-\pi,0)  \end{array} \right. \quad \Rightarrow \quad \left\{
\begin{array}{ll}
\alpha_m(x)\in {\cal P}_i(a_m),  &    a_m\in{\mathbb C}^{*},  \vspace{1mm}\\
e^{-2\pi x}\alpha_m(x)\in {\cal P}_i(a_m),  &    a_m\in{\mathbb C}^{*}.
\end{array} \right.
\ee
Just as for $N=1$, it is therefore enough to show that a
function ${\cal A}(x,p)$ given by (\ref{auxform}) and satisfying
$N$ equations of this type is uniquely determined. (Indeed, the
function ${\cal D}(x,p)$ on the lhs of (\ref{Aade}) is by
construction of the form
\be
{\cal D}(x,p)=e^{ixp}\sum_{k=0}^{N-1}d_k(x)e^{kp},\qquad d_k\in{\cal M},
\ee
and satisfies the same $N$ equations. Thus ${\cal A}-{\cal D}$ must
equal ${\cal A}$, by virtue of uniqueness.)

In order to prove uniqueness, we combine (\ref{auxform}) and
(\ref{keyeq}), obtaining
\be
\ba{l}
\ds e^{2ir_mx}\left(e^{Nr_m}+\sum_{k=0}^{N-1}c_k(x)e^{kr_m}\right)
\vspace{3mm}\\
\ds \qquad \qquad
= \alpha_m(x)\left(e^{-Nr_m}+\sum_{k=0}^{N-1}c_k(x)e^{-kr_m}\right),\qquad
m=1,\ldots,N.
\ea
\ee
These equations can be rewritten as
\be
\sum_{k=0}^{N-1}M_{mk}(x)c_k(x)=f_m(x),\qquad m=1,\ldots,N,
\ee
where
\be
M_{mk}(x)\equiv \left\{
\begin{array}{ll}
e^{2ir_mx}e^{kr_m}-\alpha_m(x)e^{-kr_m},   &  {\rm Im}\,
r_m\in(0,\pi),
\vspace{1mm}\\
e^{2i(r_m+i\pi)x}e^{kr_m}-e^{-2\pi x}\alpha_m(x)e^{-kr_m},
&  {\rm Im}\,  r_m\in(-\pi,0),
\end{array} \right.
\ee
and where the functions $f_m(x)\in{\cal M}$ need not be specified.
This system of equations has a unique solution $c(x)\in{\cal M}^N$
iff $|M(x)|\in{\cal M}^{*}$. Finally, to verify that $|M(x)|$ cannot
vanish identically, one need only take ${\rm Re}\,  x\to\infty$ to
obtain (cf.~(\ref{alm}))
\be
M_{mk}(x)\to -a_me^{-kr_m},\qquad a_m\in{\mathbb C}^{*},\quad
m=1,\ldots,N,\quad k=0,\ldots,N-1.
\ee
Since $\exp(-r_m)\ne \exp(-r_n)$ for $m\ne n$, the limit
matrix is regular, cf.~(\ref{Vdet}). Hence we must have
$|M(x)|\in{\cal M}^{*}$. \hfill \rule{3mm}{3mm}

\medskip

From the proof just given it is readily seen that we could
have allowed more general multipliers,
just as we have explicitly shown for $N=1$. But the
class of A$\Delta$Os admitting reflectionless eigenfunctions to
which we have restricted attention permits a uniform treatment
and contains all of the A$\Delta$Os of interest in later sections
and in Parts~II and~III.

\setcounter{equation}{0}
\section{Special cases: $\pbf{V_b=0}$ vs $\pbf{N_{\delta}=0}$}

In this section we are mainly concerned with features of the
potential $V_b(r,\mu;x)$ (\ref{Vbf}) and of the A$\Delta$O
$A(r,\mu)$ (\ref{Af}) for the special cases $N_{-}=0$ and
$N_{+}=0$, cf.~(\ref{NN}). Indeed, as will transpire, these two
topics are closely related.

Our first result is of a general nature: It asserts that $V_b$
cannot vanish identically for generic $r$.

\begin{theor}
Suppose $r=(r_1,\ldots,r_N)$ satisfies
\be\label{extra}
e^{r_m}\ne -e^{r_n},\qquad 1\le m<n\le N.
\ee
Then one has
\be\label{Vbnz}
V_b(r,\mu;x)\in{\cal M}^{*}.
\ee
\end{theor}

\noindent {\bf Proof.} Let us assume $V_b(x)$ vanishes
identically. Then (\ref{Vbf}) entails \be\label{Rzet} {\rm
Col}\,(R(x)-R(x-i),R(x-i)-R(x-2i),\ldots,R(x-(N-1)i)
-R(x-Ni))^t\zeta =0, \ee where ${\rm Col}\,
(\gamma_1,\ldots,\gamma_N)$ denotes the matrix with columns
$\gamma_1,\ldots,\gamma_N\in{\mathbb C}^N$. Therefore the
determinant of \be {\rm
Col}\,(R(x)-R(x-i),R(x)-R(x-2i),\ldots,R(x)-R(x-Ni)) \ee vanishes.
This remains true when we multiply the $n$th row by
$c_n\exp(-2ir_nx)$ for ${\rm Im}\,  r_n\in(0,\pi)$ and by
$c_n\exp(-2i(r_n+i\pi)x)$ for ${\rm Im}\,  r_n\in(-\pi,0)$. Due to
(\ref{R+1}) and (\ref{R+2}), the resulting matrix has ${\rm Re}\,
x\to\infty$ limit \be {\rm Row}\, (\eta_1,\ldots,\eta_N),\quad
\eta_n\equiv
\left(1-e^{2r_n},1-e^{4r_n},\ldots,1-e^{2Nr_n}\right),\quad
n=1,\ldots,N. \ee This limit matrix is of the form (\ref{Vdm}), so
it is regular iff $\exp(2r_m)\ne \exp(2r_n)$ for all pairs $m\ne
n$, cf.~Lemma~A.2. In view of our standing assumption
$\exp(r_m)\ne\exp(r_n)$, this amounts to (\ref{extra}). Therefore
we obtain a contradiction.\hfill \rule{3mm}{3mm}

\medskip

As a corollary of this general result, one infers that $V_b$
cannot vanish when either $N_{-}=0$ or $N_{+}=0$. (Indeed, in
the first/second case all of the numbers
$\exp(r_1),\ldots,\exp(r_N)$ belong to the upper/lower half
plane, so (\ref{extra}) holds true.) The following result
is of a general nature as well, but as a corollary it
yields a special case in which
$V_b$ does vanish. As a preparation we define an involution on
the space of $(r,\mu)$ by
\be\label{cC}
{\cal C} \, :\, (r,\mu) \mapsto (r^{*},\mu^{*}),
\ee
\be\label{rms}
r^{*}_n\equiv \left\{
\begin{array}{ll}
r_n-i\pi,  &  {\rm Im}\,  r_n\in(0,\pi), \vspace{1mm}\\
r_n+i\pi,  &  {\rm Im}\,  r_n\in(-\pi,0),
\end{array} \right.
\qquad \mu^{*}_n\equiv -\mu_n,\qquad n=1,\ldots,N. \ee

\begin{theor}
One has
\be\label{CR}
R(r^{*},\mu^{*};x)=-R(r,\mu;x),
\ee
\be\label{Cla}
\lambda (r^{*},\mu^{*};x)=\lambda (r,\mu;x),
\ee
\be\label{CVa}
V_a(r^{*},\mu^{*};x)=V_a(r,\mu;x),
\ee
\be\label{CVb}
V_b(r^{*},\mu^{*};x)=-V_b(r,\mu;x),
\ee
\be\label{CW}
{\cal W} (r^{*},\mu^{*};x,p)=e^{\pi x}{\cal W} (r,\mu;x,p+i\pi).
\ee
\end{theor}

\noindent
{\bf Proof.} From (\ref{defD}) and (\ref{defd}) we deduce
\be
D(r^{*},\mu^{*};x)=-D(r,\mu;x).
\ee
Likewise, (\ref{Cr}) yields
\be
C(r^{*})=-C(r).
\ee
Therefore, (\ref{CR}) follows from (\ref{sysN}). Using
(\ref{CR}) it is straightforward to check the remaining
relations, cf.~(\ref{laf}), (\ref{Vaf}), (\ref{Vbf}) and
(\ref{Wf}).\hfill \rule{3mm}{3mm}

\medskip

We proceed by observing that $V_a$, $V_b$ and ${\cal W}$ are invariant
under arbitrary permutations
\be\label{perm}
\pi \, :\, (r,\mu)\mapsto (\pi(r),\pi(\mu)),\qquad \pi \in S_N.
\ee
In view of (\ref{CVb}), we therefore have the implication
\be
(r^{*},\mu^{*})=(\pi(r),\pi(\mu)),\qquad \pi\in S_N  \Rightarrow
V_b(r,\mu;x)=0.
\ee
We conjecture that this sufficient condition is also necessary:
\be\label{expect}
V_b(r,\mu;x)=0 \Rightarrow
(r^{*},\mu^{*})=(\pi(r),\pi(\mu)),\qquad \pi\in S_N .\qquad \quad
 (?)
\ee From substantial later developments (cf.~(\ref{Vbnzero})
below), we will be able to conclude that the sufficient condition
(\ref{extra}) for $V_b\ne 0$ can be improved to $r^{*}\ne
\pi(r),\forall \, \pi\in S_N$, but a proof of our expectation
(\ref{expect}) has not materialized thus far.

In view of permutation  invariance, we are free to choose a
convenient ordering for the special case just considered. Since
the involution ${\cal C}$ (\ref{cC}) maps the set $\{
\exp(r_1),\ldots$, $\exp(r_N)\}$ to $\{
-\exp(r_1),\ldots,-\exp(r_N)\}$, this case can only arise when
$N=2M$, $M\in{\mathbb N}$, and $N_{+}=N_{-}=M$. Choosing
$r_1,\ldots,r_M$ in the strip $\{ {\rm Im}\,  r\in(0,\pi)\}$, we
may choose $r_{M+n}$ equal to $r_n-i\pi$. Setting \be\label{piM}
\pi_M\equiv \left( \begin{array}{cc}
0  &  {\bf 1}_M  \\
{\bf 1}_M  &  0  \end{array} \right),
\ee
we then have
\be\label{rmus}
r^{*}=\pi_M r,\qquad
\mu^{*}=\pi_M \mu,
\ee
\be\label{Vb0}
V_b((r_1,\ldots,r_M,r_1-i\pi,\ldots,r_M-i\pi,
\mu_1,\ldots,\mu_M,-\mu_1,\ldots,-\mu_M;x)=0,
\ee
where ${\rm Im}\,
r_n\in(0,\pi)$ for $ n=1,\ldots,M$.

We proceed by exploiting the above insights to obtain some
striking properties of the special cases $N_{\delta}=0$, $\delta =+,-$.
First, we introduce the ``square root'' A$\Delta$Os
\be\label{Spm}
S_{\pm}(r,\mu)\equiv T_{i/2}\pm V(r,\mu;x)T_{-i/2},
\ee
where
\be\label{defV}
V(r,\mu;x)\equiv \lambda(r,\mu;x)/\lambda(r,\mu;x+i/2).
\ee
From (\ref{Vaf}) we then deduce
\be\label{S2}
S_{\delta}(r,\mu)^2=T_i+V_a(r,\mu;x)T_{-i}+\delta V_s(r,\mu;x),\qquad
\delta =+,-,
\ee
where we have set
\be\label{Vs}
V_s(r,\mu;x)\equiv V(r,\mu;x-i/2)+V(r,\mu;x).
\ee
It is understood here that $r$ is arbitrary. But only when
$N_{\delta}=0$ we can show that $S_{-\delta}(r,\mu)$ has a special
significance, as detailed in our next theorem.

\begin{theor}
Fix $N$ distinct numbers $r_1,\ldots,r_N$ in the strip $\{ {\rm Im}\,
r\in(0,\pi)\}$ and multipliers
$\mu_1(x),\ldots,\mu_N(x)$ satisfying (\ref{req3}), and set
\be
r^{+}\equiv (r_1,\ldots,r_N),\qquad r^{-}\equiv
(r_1-i\pi,\ldots,r_N-i\pi).
\ee
Then one has the A$\Delta$Es
\be\label{Sade}
S_{\delta}\left(r^{\delta}, \mu\right){\cal W}(r^{\delta}, \mu
;x,p)=\left(e^{p/2}+\delta e^{-p/2}\right){\cal
W}\left(r^{\delta},\mu;x,p\right),
\qquad \delta=+,-,
\ee
and identities
\be\label{VV}
V_s\left(r^{+},\mu;x\right)=V_b\left(r^{+},\mu;x\right)+2,
\ee
\be\label{VR}
V\left(r^{+},\mu;x\right)=\sum_{n=1}^N\left(R_n\left(r^{+},\mu;x-i/2\right)-
R_n\left(r^{+},\mu;x\right)\right)+1.
\ee
Moreover, introducing
\be\label{rs}
r_s\equiv (r_1/2,\ldots,r_N/2,r_1/2-i\pi,\ldots,r_N/2-i\pi),
\ee
\be\label{mur}
\mu_r(x)\equiv
\left(2e^{-r_1/2}\mu_1(x/2),\ldots,2e^{-r_N/2}\mu_N(x/2)\right),
\ee
one has the relations
\be\label{RR}
2e^{-r_n/2}R_n(r_s,\mu_r,-\mu_r;2x)=R_n\left(r^{+},\mu;x\right),
\qquad n=1,\ldots,N,
\ee
\be\label{WW}
{\cal W}(r_s,\mu_r,-\mu_r;2x,p/2)={\cal W}\left(r^{+},\mu;x,p\right),
\ee
\be\label{lala}
\lambda (r_s,\mu_r,-\mu_r;2x)=\lambda \left(r^{+},\mu;x\right),
\ee
\be\label{VaV}
V_a(r_s,\mu_r,-\mu_r;2x)=V\left(r^{+},\mu;x\right),
\ee
\be\label{Vbz}
V_b(r_s,\mu_r,-\mu_r;2x)=0.
\ee
\end{theor}

\noindent
{\bf Proof.} We begin by proving (\ref{RR}). To
this end we observe
\be
D(r_s,\mu_r,-\mu_r;2x)=\left( \begin{array}{rr}
E  &  0  \\
0  &  -E \end{array}\right) ,\qquad
E\equiv D\left(r^{+}/2,\mu_r;2x\right),
\ee
cf.~(\ref{defD}), (\ref{defd}). Similarly, we have
\be
C(r_s)=\left( \begin{array}{rr}
A  &  B  \\
-B  &  -A \end{array}\right) ,
\ee
with
\be\label{AB}
A_{mn}\equiv \left(e^{r_m/2}-e^{-r_n/2}\right)^{-1}\!,\quad B_{mn}\equiv
\left(e^{r_m/2}+e^{-r_n/2}\right)^{-1}\!,\quad m,n=1,\ldots,N,\!\!\!\!
\ee
cf.~(\ref{Cr}). Using obvious notation, the system
(\ref{sysN}) with $r\to r_s$, $\mu\to (\mu_r,-\mu_r)$ can
therefore be written as
\be\label{sys2N}
\left( \left( \begin{array}{rr}
E  &  0  \\
0  &  -E \end{array}\right) +
\left( \begin{array}{rr}
A  &  B  \\
-B  &  -A \end{array}\right) \right)
\left( \begin{array}{l}
R^{+}  \\
R^{-} \end{array}\right) =
\left( \begin{array}{l}
\zeta^{+}  \\
\zeta^{-} \end{array}\right) .
\ee
Now when we multiply this by $\pi_N$ (given by (\ref{piM})
with $M\to N$), we obtain
\be\label{sys2Nnew}
\left( \left( \begin{array}{rr}
E  &  0  \\
0  &  -E \end{array}\right) +
\left( \begin{array}{rr}
A  &  B  \\
-B  &  -A \end{array}\right) \right)
\left( \begin{array}{l}
-R^{-}  \\
-R^{+} \end{array}\right) =
\left( \begin{array}{l}
\zeta^{+}  \\
\zeta^{-} \end{array}\right) .
\ee
Comparing (\ref{sys2N}) and (\ref{sys2Nnew}), we deduce
\be\label{R+-}
R^{-}=-R^{+}.
\ee
Substituting this in (\ref{sys2N}), the first $N$ equations
become
\be\label{sysred}
(E+A-B)R^{+}=\zeta^{+}.
\ee

Next, we use (\ref{AB}) to get
\be
A_{mn}-B_{mn}=2e^{-r_n/2}\left(e^{r_m}-e^{-r_n}\right)^{-1},\qquad
m,n=1\ldots,N.
\ee
Furthermore, we have from (\ref{mur}) and (\ref{defd})
\be
d\left(r_n/2,\mu_{r,n};2x\right)=2e^{-r_n/2}d\left(r_n,\mu_n;x\right),\qquad
n=1,\ldots,N.
\ee
Hence (\ref{sysred}) entails (\ref{RR}).

Using (\ref{R+-}) and (\ref{RR}), we now obtain
\be
\ba{l}
\ds {\cal W}(r_s,\mu_r,-\mu_r;2x,p/2)    =    e^{ixp}\!\left(\!
1-\!\sum_{n=1}^N\left(\!
\frac{R_n(r_s,\mu_r,-\mu_r;2x)}{e^{p/2}-e^{-r_n/2}}-
\frac{R_n(r_s,\mu_r,-\mu_r;2x)}{e^{p/2}+e^{-r_n/2}}
\!\right) \!\right)
\vspace{3mm} \\
\ds  \phantom{{\cal W}(r_s,\mu_r,-\mu_r;2x,p/2) }
=   e^{ixp}\left(
1-\sum_{n=1}^N\left(e^p-e^{-r_n}\right)^{-1}R_n\left(r^{+},\mu;x\right)\right)
\vspace{3mm}\\
\ds   \phantom{{\cal W}(r_s,\mu_r,-\mu_r;2x,p/2) } =   {\cal
W}\left(r^{+},\mu;x,p\right), \ea\hspace{-20mm} \ee which proves
(\ref{WW}). Next, we combine (\ref{laf}), (\ref{RR}) and
(\ref{rs}) to obtain (\ref{lala}). Then (\ref{VaV}) follows from
(\ref{Vaf}), (\ref{lala}) and (\ref{defV}). Moreover, recalling
(\ref{Vb0}), we deduce (\ref{Vbz}). (Alternatively, (\ref{Vbz})
can be inferred directly from (\ref{R+-}) and (\ref{Vbf}).)

Consider now the A$\Delta$E
\be
\ba{l}
\ds {\cal W}(r_s,\mu_r,-\mu_r;2x-i,p/2)+V_a(r_s,\mu_r,-\mu_r;2x)
{\cal W}(r_s,\mu_r,-\mu_r;2x+i,p/2)
\vspace{2mm}\\
\ds \qquad \qquad =\left(e^{p/2}+e^{-p/2}\right){\cal
W}(r_s,\mu_r,-\mu_r;2x,p/2),
\ea
\ee
satisfied by the lhs of (\ref{WW}) (due to (\ref{Vbz})).
Rewriting it for the rhs of (\ref{WW}) and using (\ref{VaV}),
this yields (\ref{Sade}) with $\delta=+$.

To arrive at (\ref{Sade}) with $\delta=-$, we invoke
Theorem~3.2. Specifically, from (\ref{Cla}) and (\ref{defV})
we deduce
\be\label{pm1}
V\left(r^{-},-\mu;x\right)=V\left(r^{+},\mu;x\right),
\ee
and from (\ref{CW}) we have
\be\label{pm2}
{\cal W}\left(r^{-},-\mu;x,p\right)=e^{\pi x}{\cal
W}\left(r^{+},\mu;x,p+i\pi\right).
\ee
Therefore, the above A$\Delta$E (\ref{Sade}) with $\delta=-$
readily follows from its $\delta=+$ counterpart.

We continue by proving (\ref{VV}). On account of (\ref{S2})
we have
\be
S_{+}\left(r^{+},\mu\right)^2-A\left(r^{+},\mu\right)-2
=V_s\left(r^{+},\mu;x\right)-V_b\left(r^{+},\mu;x\right)-2.
\ee
Since $S_{+}^2$ has eigenvalue $(\exp(p/2)+\exp(-p/2))^2$ and
$A$ has eigenvalue $\exp(p)+\exp(-p)$ on
${\cal W}\left(r^{+},\mu;x,p\right)$, the A$\Delta$O difference on the lhs
annihilates ${\cal W}\left(r^{+},\mu;x,p\right)$. Hence the rhs vanishes,
which amounts to (\ref{VV}).

Finally, we prove (\ref{VR}). To this end we use the A$\Delta$E
(\ref{Sade}) with $\delta=+$ for the auxiliary wave function
(\ref{defaux}). It entails
\be\label{newade}
\ba{l}
\ds \left(e^p+1\right)\left[e^{Np}+\sum_{k=0}^{N-1}c_k(x-i/2)e^{kp}\right]
+\left(1+e^{-p}\right)
V(x)\left[e^{Np}+
\sum_{k=0}^{N-1}c_k(x+i/2)e^{kp}\right]
\vspace{4mm}\\
\ds \qquad  -\left(e^{p/2}+e^{-p/2}\right)^2
\left[e^{Np}+\sum_{k=0}^{N-1}c_k(x)e^{kp}\right] = 0.
\ea\hspace{-15mm}
\ee
The vanishing of the coefficient of $\exp(Np)$ now amounts to
(\ref{VR}), cf.~(\ref{cb}).\hfill \rule{3mm}{3mm}

\strut\hfill

We would like to add that we have not found a direct proof of
the main results (\ref{Sade})--(\ref{VR}) of this theorem.
(That is, a proof that avoids the detour via the doubled-up
system involving (\ref{rs}), (\ref{mur}).)

In this connection we point
out that (\ref{Sade})--(\ref{VR}) would also follow from the
identity
\be\label{Rid}
\sum_{n=1}^N\left(R_n\left(r^{+},\mu;x-i/2\right)-R_n\left(r^{+},
\mu;x\right)\right)+1=
\frac{1+\sum\limits_{n=1}^Ne^{r_n}R_n\left(r^{+},\mu;x\right)}
{1+\sum\limits_{n=1}^Ne^{r_n}R_n\left(r^{+},\mu;x+i/2\right)}.\hspace{-5mm}
\ee
To see this, note first that this identity is equivalent to
(\ref{VR}), cf.~(\ref{defV}). Now assume one could prove
(\ref{Rid}) directly. Then one easily obtains (\ref{VV}) by
using (\ref{Vs}) and (\ref{Vbf}). Moreover, the A$\Delta$E
(\ref{Sade}) with
$\delta=+$ follows from the uniqueness argument in the proof of
Theorem~2.1, with the lhs of (\ref{newade}) playing the role
of the lhs of (\ref{Aade}). Finally, the $\delta=-$ counterpart
follows as before from Theorem~3.2, cf.~(\ref{pm1}),
(\ref{pm2}).

\setcounter{equation}{0}
\section{The injectivity problem}

In this section we return to the general setting of Theorem~2.1.
Thus we start from comp\-lex numbers $r_1,\ldots,r_N$ satisfying
(\ref{req1}) and (\ref{req2}), and multipliers
$\mu_1,\ldots,\mu_N$ satisfying (\ref{req3}). For such $(r,\mu)$
we have defined a wave function ${\cal W}(x,p)$ via (\ref{Wf}) and
an A$\Delta$O $A$ via (\ref{Af}), with $A$ having eigenvalue
$\exp(p)+\exp(-p)$ on ${\cal W}(x,p)$. As already observed
(cf.~(\ref{perm})), we obtain the same ${\cal W}(x,p)$ and $A$
when we permute $r$ and $\mu$. Let us denote the above space of
$(r,\mu)$ with $N$ varying over ${\mathbb N}$, divided by the
action of the symmetric groups $S_N,N\in {\mathbb N}$, by ${\cal
D}_{\rm IST}$. Then we have well-defined maps \be\label{phiw}
\Phi_{{\cal W}}\, :\, {\cal D}_{\rm IST}\to \{ {\rm wave\
functions}\} ,\qquad (r,\mu)\mapsto {\cal W}(r,\mu;x,p), \ee
\be\label{phia} \Phi_{A}\, :\, {\cal D}_{\rm IST}\to \{ {\rm
A}\Delta{\rm Os}\} ,\qquad
 (r,\mu)\mapsto A(r,\mu).
\ee

Within the algebraic framework of this paper, a natural question
now arises. It reads: Are the above ``generalized IST'' maps
$\Phi_{{\cal W}}$ and $\Phi_A$ one-to-one? It is not hard to see
that the answer is ``yes'' for $\Phi_{{\cal W}}$, cf.~Theorem~4.1.
We conjecture that the answer is ``yes'' for~$\Phi_A$ as well, and
we go a long way towards proving this, cf.~Lemmas~4.2--4.4.

Before embarking on the details, let us point out that the
injectivity conjecture just made is closely related to our
expectation (\ref{expect}). Indeed, if (\ref{expect}) is
false, then there exists $(r_0,\mu_0)$ such that
$V_b(r_0,\mu_0;x)=0$, yet $\left(r_0^{*},\mu_0^{*}\right)$ is not obtained
from $(r_0,\mu_0)$ via a permutation. Now from (\ref{CVa}) and
(\ref{CVb}) we have $A(r_0,\mu_0)=A\left(r_0^{*},\mu_0^{*}\right)$, so
$\Phi_A$ would not be 1-1. As a consequence, our injectivity
conjecture is stronger than (\ref{expect}).

\begin{theor}
The map $\Phi_{{\cal W}}$ (\ref{phiw}) is an injection.
\end{theor}

\noindent
{\bf Proof.} Assuming ${\cal W}(r_0,\mu_0;x,p)$ equals
${\cal W}(r,\mu;x,p)$ for some $(r,\mu)$, we compare the poles of
these wave functions (given by (\ref{Wf})) in the strips ${\rm Im}\,
p\in(0,\pi)$ and ${\rm Im}\,  p\in(-\pi,0)$ to deduce that $N=N_0$
and that $r$ is related to $r_0$ by a permutation. Reordering,
we may as well assume $r=r_0$. Then the residue vectors
$R(r_0,\mu_0;x)$ and $R(r_0,\mu;x)$ coincide. Using
(\ref{sysN}), we now obtain
\be
[D(r_0,\mu_0;x)-D(r_0,\mu;x)]R(x)=0.
\ee
Since $R\in{\cal M}^{*N}$, this yields $\mu =\mu_0$,
cf.~(\ref{defD}), (\ref{defd}).\hfill \rule{3mm}{3mm}

\strut\hfill

We now turn to $\Phi_A$ (\ref{phia}). We begin by pointing out
that it is already quite unclear whether for a given
${\cal W}(x,p)\equiv {\cal W}(r_0,\mu_0;x,p)$ there might not be
infinitely many $\tilde{{\cal W}}(x,p)$ of the form
${\cal W}(r,\mu;x,p)$ that are related to ${\cal W}(x,p)$ via
(\ref{Wamb}). Indeed, whenever $(r,\mu)$ has this property,
(\ref{Wamb}) implies $A(r,\mu)=A(r_0,\mu_0)$. (To check the
asserted implication, note that $i$-periodicity
of $\nu_1(x),\ldots,\nu_M(x)$ entails
$\tilde{{\cal W}}(x,p)$ is an $A(r_0,\mu_0)$-eigenfunction with
eigenvalue $\exp(p)+\exp(-p)$, and recall the conclusion below
(\ref{Vac}).) In the next lemma we exclude in particular such
ambiguities.

\begin{lem}
Suppose that for a given $(r_0,\mu_0)$ there exists $(r,\mu)$
such that
\be\label{AAt}
A(r,\mu)=A(r_0,\mu_0).
\ee
Then one has $N=N_0$ and $r$ equals $\pi(r_0)$ with $\pi$ a
permutation. Moreover, for all $p\in{\mathbb C}$ the auxiliary wave
functions are related by
\be\label{nup}
{\cal A}(r,\mu;x,p)=\nu_{+}(x,p){\cal A}(r_0,\mu_0;x,p),\qquad
\nu_{+}(\cdot,p)\in{\cal P}_i.
\ee
Finally, for all $(r,\mu)\in {\cal D}_{\rm IST}$ one has the
implication
\be\label{Vbnzero}
r^{*}\ne \pi(r),\qquad \forall\, \pi\in S_N \Rightarrow V_b(r,\mu;x)\ne 0.
\ee
\end{lem}

\noindent
{\bf Proof.} Our reasoning makes extensive use of Casorati
determinants. In Appendix~B we have summarized the pertinent
general features. Moreover, we have explicitly determined some
relevant Casorati determinants. As a consequence, we deduce
that the auxiliary wave functions ${\cal A}(r_0,\mu_0;x,\pm p)$
yield a basis (over ${\cal P}_i$) for the space of all meromorphic
solutions to the A$\Delta$E
\be\label{Aade2}
F(x-i)+V_a(r_0,\mu_0;x)F(x+i)+
\left[V_b(r_0,\mu_0;x)-e^p-e^{-p}\right]F(x)=0,
\ee
provided
\be\label{pcond}
\exp (p)\ne \pm 1,\exp(\pm r_{0,1}),\ldots, \exp(\pm
r_{0,N_0}).
\ee

Now the assumption (\ref{AAt}) entails that the functions
${\cal A}(r,\mu;x,\pm p)$ solve (\ref{Aade2}) as well. With
(\ref{pcond}) in force, we then have
\be\label{Alc}
{\cal A}(r,\mu;x,p)=\nu_{+}(x,p){\cal A}(r_0,\mu_0;x,p)+
\nu_{-}(x,p){\cal A}(r_0,\mu_0;x,-p).
\ee
Here, the $i$-periodic multipliers are quotients of Casorati
determinants, cf.~(\ref{F3}), (\ref{nuj}). We claim that
$\nu_{-}(x,p)$ vanishes.

To prove this claim, we need only show that
\be
D(x,p)\equiv {\cal D}({\cal A}(r,\mu;x,p),{\cal A}(r_0,\mu_0;x,p))
\ee
is zero. To this end we recall $D(x,p)$ satisfies the A$\Delta$E
\be\label{DV}
D(x,p)=V_a(x)D(x+i,p),
\ee
cf.~(\ref{casade}). Now $D(x,p)$ is of the form
\be
D(x,p)=e^{2ixp}e^p\sum_{l=0}^{N_0+N-1}e^{lp}d_l(x),
\ee
cf.~(\ref{casdef}) and (\ref{auxform}). Therefore, (\ref{DV})
yields
\be
\sum_{l=0}^{N_0+N-1}e^{(l+1)p}d_l(x)=
\sum_{l=0}^{N_0+N-1}e^{(l-1)p}d_l(x+i)V_a(x).
\ee
Thus we obtain recursively $d_l(x)=0,l=N_0+N-1,\ldots,0$,
which proves our claim.

As a consequence, (\ref{nup}) holds true for all $p$
satisfying (\ref{pcond}).
Since the auxiliary wave functions are entire in $p$ and
belong to ${\cal M}^{*}$ for all $p\in{\mathbb C}$, (\ref{nup}) holds for
arbitrary $p$.

Now (\ref{nup}) yields in particular
\be
{\cal A}(r,\mu;x,\pm r_n)=\nu_{+}(x,\pm r_n){\cal A}(r_0,\mu_0;x,\pm
r_n),
\ee
and we also have
\be
{\cal A}(r,\mu;x,r_n)=\alpha_n(x){\cal A}(r,\mu;x,-r_n),
\ee
with $\alpha_n$ $i$-periodic, cf.~(\ref{keyeq}), (\ref{alm}).
Hence,
we get
\be
{\cal
A}(r_0,\mu_0;x,r_n)=\alpha_n(x)\frac{\nu_{+}(x,-r_n)}{\nu_{+}(x,r_n)}{\cal
A}(r_0,
\mu_0;x,-r_n),
\ee
so ${\cal A}(r_0,\mu_0;x,r_n)/{\cal A}(r_0,\mu_0;x,-r_n)$ is
$i$-periodic. But then the Casorati determinant of
${\cal A}(r_0,\mu_0;x,r_n)$ and ${\cal A}(r_0,\mu_0;x,-r_n)$ vanishes.
Since $\exp(r_n)\ne \pm 1$, this implies that $r_n$ or $-r_n$
belongs to $\{ r_{0,1},\ldots,r_{0,N_0}\} $, on account of
(\ref{casA}). Likewise, we deduce that $r_{0,k}$ or $-r_{0,k}$
belongs to $\{ r_1,\ldots,r_N\}$. Clearly, this entails $N=N_0$
and the existence of a permutation $\pi\in S_N$ such that
\be\label{rr0}
r_n=s_n\pi(r_0)_n,\qquad s_n\in\{ -1,1\} ,\qquad n=1,\ldots,N.
\ee

Next, we show that all of the signs $s_n$ are positive.
Indeed, let us assume $s_k=-1$, so as to derive a
contradiction. Reordering, we may and will choose $\pi ={\rm
id}$ in (\ref{rr0}). Now from (\ref{sysN}) we deduce
\be\label{Rk}
{\cal W}(r,\mu;r_k)=e^{ir_kx}d(r_k,\mu_k;x)R_k(r,\mu;x).
\ee
Likewise, our assumption entails
\be\label{Rkh}
{\cal W}(r_0,\mu_0;-r_k)=e^{-ir_kx}d(-r_k,\mu_{0,k};x)
R_k(r_0,\mu_0;x).
\ee
Recalling the definition (\ref{defaux}) of the auxiliary wave
function and the relations (\ref{keyeq}), it readily follows
from (\ref{nup}) that the quotient of the functions (\ref{Rk})
and (\ref{Rkh}) is $i$-periodic. Using the definition
(\ref{defd}) of $d$, we then deduce
\be
R_k(r,\mu;x)/R_k(r_0,\mu_0;x)=e^{2ir_kx}\nu(x),\qquad
\nu\in{\cal P}_i.
\ee

From this equation we now obtain the desired contradiction.
Indeed, the ${\rm Re}\,  x\to -\infty$ limit of the quotient on the lhs
is a non-zero constant $q$, due to (\ref{R-}). Hence we obtain
\be
\lim_{{\rm Re}\, x\to -\infty} e^{2ir_kx}\nu(x)=q\ne 0.
\ee
But since $\nu(x)$ is $i$-periodic, we also have
\be
q=\lim_{{\rm Re}\, x\to -\infty} e^{2ir_k(x-i)}\nu(x-i)=e^{2r_k}q.
\ee
As $\exp(2r_k)\ne 1$, this entails $q=0$, a contradiction.

The implication (\ref{Vbnzero}) is now a simple corollary: If
one has $V_b=0$, yet $r^{*}\ne \pi(r)$ for all $\pi\in S_N$,
then one gets $A(r,\mu)=A(r^{*},\mu^{*})$ from (\ref{CVa}) and
(\ref{CVb}), which contradicts the first assertion of the
lemma.\hfill \rule{3mm}{3mm}

\medskip

The remaining problem is to show that the assumption
\be\label{ass2}
A(r,\mu)=A(r,\hat{\mu})
\ee
entails $\hat{\mu}=\mu$. We believe this is true, but have no
complete proof. From (\ref{ass2}) and (\ref{nup}) it is
obvious that the two pertinent wave functions
\be\label{W1}
{\cal W}(r,\mu;x,p)\equiv e^{ixp}\left( 1-\sum_{n=1}^N
\frac{R_n(x)}{e^p-e^{-r_n}}
\right),
\ee
\be\label{W2}
{\cal W}(r,\hat{\mu};x,p)\equiv e^{ixp}\left( 1-\sum_{n=1}^N
\frac{\hat{R}_n(x)}{e^p-e^{-r_n}}
\right),
\ee
are related by an $i$-periodic multiplier. This is equivalent
to the identity
\be
\ba{l}
\ds \left( 1-\sum_{m=1}^N
\frac{R_m(x)}{e^p-e^{-r_m}}
\right)
\left( 1-\sum_{n=1}^N
\frac{\hat{R}_n(x+i)}{e^p-e^{-r_n}}
\right)
\vspace{3mm}\\
\ds \qquad \qquad = \left( 1-\sum_{m=1}^N
\frac{\hat{R}_m(x)}{e^p-e^{-r_m}}
\right)
\left( 1-\sum_{n=1}^N
\frac{R_n(x+i)}{e^p-e^{-r_n}}
\right).
\ea
\ee
Multiplying by $(\exp(p)-\exp(-r_k))^2$ and letting $p\to
-r_k$, we get
\be
R_k(x)\hat{R}_k(x+i)=\hat{R}_k(x)R_k(x+i).
\ee
From this we deduce
\be\label{RtR}
\hat{R}_k(x)=\nu_k(x)R_k(x),\qquad \nu_k\in{\cal P}_i,\qquad
k=1,\ldots,N.
\ee

Taking now the difference of the A$\Delta$Es satisfied by
(\ref{W1}) and (\ref{W2}), we obtain the identity
\be
e^pH(x-i,p)+e^{-p}V_a(x)H(x+i,p)+\left[V_b(x)
-e^p-e^{-p}\right]H(x,p)=0,
\ee
where $V_a$, $V_b$ denote the potentials in
$A(r,\mu)=A(r,\hat{\mu})$, and where we have introduced
\be
H(x,p)\equiv
\sum_{n=1}^N\frac{1}{e^{p}-e^{-r_n}}[1-\nu_n(x)]
R_n(x).
\ee
Clearly, the vanishing residues for
$p=-r_k$ yield no new information. But the limit for ${\rm Re}\,
p\to \infty$ yields
\be\label{conseq}
\sum_{n=1}^N[1-\nu_n(x)]
[R_n(x-i)-R_n(x)]=0.
\ee
We are now going to use the consequence (\ref{conseq}) of
(\ref{ass2}) to prove our conjecture $\hat{\mu}=\mu$ under an
additional hypothesis, which is however generically satisfied.

\begin{lem}
Suppose $N_{+}$ or $N_{-}$ vanishes. Then one has the implication
\be\label{impl} A(r,\mu)=A(r,\hat{\mu})\quad \Rightarrow\quad
\hat{\mu}=\mu. \ee More generally, suppose $r=(r_1,\ldots,r_N)$
satisfies \be\label{extra2} e^{r_m}\ne -e^{r_n},\qquad 1\le m<n\le
N. \ee Then (\ref{impl}) holds true as well.
\end{lem}

\noindent
{\bf Proof.} We need only show that (\ref{extra2}) entails
(\ref{impl}). To this end we exploit
(\ref{conseq}). Since $\nu_n(x)$
is $i$-periodic, it entails that (\ref{Rzet}) holds true, with
$\zeta$ replaced by $h(x)\equiv
(1-\nu_1(x),\ldots,1-\nu_N(x))^t$. Just as in the proof of
Theorem~3.1, a contradiction now arises when $h(x)$ does not
vanish. Thus we must have $\nu_n(x)=1$, $n=1,\ldots,N$. But then
the two wave functions (\ref{W1}), (\ref{W2}) are equal, so
that Theorem~4.1 yields $\hat{\mu}=\mu$.
\hfill \rule{3mm}{3mm}

\medskip

Another approach to the injectivity question that may be
useful in further studies results from our last lemma in this
section. Specifically, we assume that (\ref{ass2}) holds true,
so that (\ref{RtR}) follows. Now we introduce the $i$-periodic
matrices
\be
\nu(x)\equiv {\rm diag}\,  (\nu_1(x),\ldots,\nu_N(x)),
\ee
\be
Q(x)\equiv D(r,\mu;x)D(r,\hat{\mu};x)^{-1}={\rm diag}\,
(\mu_1(x)/\hat{\mu}_1(x),\ldots,\mu_N(x)/\hat{\mu}_N(x)),
\ee
\be
P(x)\equiv C(r)-Q(x)\nu(x)^{-1}C(r)\nu(x).
\ee

\begin{lem}
One has
\be\label{PR}
P(x)(R(x)-R(x-i))=0.
\ee
In particular, $P(x)$ has vanishing determinant.
\end{lem}

\noindent
{\bf Proof.} From (\ref{sysN}) we obtain
\be\label{sd1}
D(x)R(x)-D(x-i)R(x-i)+C(R(x)-R(x-i))=0.
\ee
Next, we exploit the $i$-periodicity of $\nu(x)$ to rewrite
the analog of (\ref{sd1}) for $\hat{R}(x)=\nu(x)R(x)$ as
\be\label{sd2}
D(x)R(x)-D(x-i)R(x-i)+Q(x)\nu(x)^{-1}C\nu(x)(R(x)-R(x-i))=0.
\ee
Subtracting (\ref{sd2}) from (\ref{sd1}), we get
\be
P(x)(R(x)-R(x-i))=0.
\ee
Now from (\ref{sysN}) and (\ref{defD}) it is evident that
$R(x)$ is not $i$-periodic. Therefore the lemma follows.
\hfill \rule{3mm}{3mm}

\medskip

Obviously, the expected equality $\hat{\mu}=\mu$ is equivalent
to all diagonal elements of $P(x)$ being zero. Introducing the
determinant
\be\label{defcD}
{\cal D}(r,\mu;x)\equiv
|{\rm Col}\,
(R(r,\mu;x)-R(r,\mu;x-i),\ldots,R(r,\mu;x)-R(r,\mu;x-Ni))|,
\ee
one readily sees that
$P(x)$ actually vanishes identically when ${\cal D}(x)\in{\cal M}^{*}$.
Unfortunately, the ${\rm Re}\,  x\to \infty$
asymptotics of the matrix  on the rhs of (\ref{defcD}) leads
for the third time to the condition
(\ref{extra2}), cf.~the proof of Theorem~3.1.

In this connection it should be pointed out that
${\cal D}(x)$ does not alway belong to ${\cal M}^{*}$.
 Indeed, the arguments in the
proof of Theorem~3.3 leading to the relation $R^{-}=-R^{+}$
(\ref{R+-}) give rise to the implication
\be\label{impl2}
\left(r^{*},\mu^{*}\right)=(\pi(r),\pi(\mu)),\pi\in S_N \Rightarrow
{\cal D}(r,\mu;x)=0.
\ee
(Note that (\ref{rs}) does not yield all of the $r$ allowed in
(\ref{impl2}), since the imaginary parts of the numbers
$r_{s,n}$ are all in $(-\pi,-\pi/2)\cup(0,\pi/2)$. But this
restriction is not used to arrive at (\ref{R+-}).)

It may well be that the converse of (\ref{impl2}) is also
valid. It would be worthwile to study this issue further, since
it clearly has a bearing on the above-mentioned conjectures. It
is also connected to another natural question we leave open,
namely: Do there exist $(r,\mu)\in {\cal D}_{\rm IST}$ such that
$V_a(r,\mu;x)=1$? Indeed, from (\ref{Vaf}) and (\ref{laf}) one
sees that this is equivalent to vanishing of
\be
\sum_{n=1}^N e^{r_n} (R_n(r,\mu;x)-R_n(r,\mu;x-i)).
\ee
Therefore, ${\cal D}(r,\mu;x)\in{\cal M}^{*}$ implies
$V_a(r,\mu;x)-1\in{\cal M}^{*}$. (In particular, one has $V_a(x)\ne 1$
when (\ref{extra2}) holds true.)

\renewcommand{\thesection}{A}
\setcounter{equation}{0}
\setcounter{lem}{0}


\section*{Appendix A.  Cauchy and
Vandermonde matrices}

As is well known, the Cauchy matrix
\be
C_{ij}\equiv \frac{1}{x_i-y_j},\qquad i,j=1,\ldots,N,
\ee
with $x_1,\ldots,x_N,y_1,\ldots,y_N$ distinct complex numbers,
has determinant
\be
|C|=\prod_i\frac{1}{x_i-y_i}\prod_{i<j}
\frac{(x_i-x_j)(y_i-y_j)}{(x_i-y_j)(y_i-x_j)}.
\ee
This Cauchy identity shows in particular that $C$ is regular.
In the main text we work with the matrix $C(r)$ obtained by
substituting
\be
x_i=e^{r_i},\qquad y_j=e^{-r_j},
\ee
with
\be
{\rm Im}\,  r_k\in (0,\pi)\cup(-\pi,0),\quad k=1,\ldots,N,\quad e^{r_m}\ne
e^{\pm r_n},\quad  1\le m<n\le N.
\ee

Basically, we already obtained
the following identities in
Lemma~A.7 of Ref.~\cite{aa3}, where we employed a slightly
different Cauchy matrix. For completeness we include a proof.

\begin{lem}
One has the identities \be\label{Crid} \ba{l} \ds
\left(C(r)^{-1}\zeta\right)_m=\left(e^{r_m}-e^{-r_m}\right)\prod_{n=1,\,n\ne
m}^N \frac{e^{-r_m}-e^{r_n}}{e^{-r_m}-e^{-r_n}},
\vspace{3mm}\\
\ds  \zeta\equiv
(1,\ldots,1)^t,\quad m=1,\ldots,N.
\ea
\ee
\end{lem}

\noindent {\bf Proof.} This is equivalent to the functional
identities \be\label{fid}
\sum_{k=1}^N\frac{1}{e^{r_j}-e^{-r_k}}\left(e^{r_k}-e^{-r_k}\right)
\prod_{l=1,\,l\ne
k}^N\frac{e^{-r_k}-e^{r_l}}{e^{-r_k}-e^{-r_l}}=1,\qquad
 j=1,\ldots,N.
\ee By permutation invariance we need only prove (\ref{fid}) for
$j=1$. In that case the lhs can be rewritten as $F(r_1)$, with
\be\label{Frho} F(\rho)\equiv
\prod_{l=2}^N\frac{e^{-\rho}-e^{r_l}}{e^{-\rho}-e^{-r_l}}+\sum_{k=2}^N
\frac{e^{r_k}-e^{-r_k}}{e^{-\rho}-e^{-r_k}}\prod_{l=2,\, l\ne k}^N
\frac{e^{-r_k}-e^{r_l}}{e^{-r_k}-e^{-r_l}}. \ee Clearly, $F(\rho)$
is $2\pi i$-periodic and has limit 1 for ${\rm Re}\, \rho\to
-\infty$. Since it has a finite limit for ${\rm Re}\,
\rho\to\infty$ as well, it suffices to verify that the residues at
the simple poles $r_2,\ldots,r_N$ cancel, which is routine. \hfill
\rule{3mm}{3mm}

\medskip

In Appendix~C we have occasion to use the identity
\be\label{Csum}
\sum_{m=1}^N \left(C(r)^{-1}\zeta\right)_m=\sum_{m=1}^N
\left(e^{r_m}-e^{-r_m}\right),
\ee
which can be derived from (\ref{Crid}) and its proof. Indeed,
substituting (\ref{Crid}) on the lhs, and setting $r_1=\rho$,
we see that (\ref{Csum}) amounts to the function
\be
\ba{l}
\ds G(\rho)    \equiv   \left(e^{\rho}-e^{-\rho}\right)\left(
\prod_{l=2}^N\frac{e^{-\rho}-e^{r_l}}{e^{-\rho}-e^{-r_l}}-1
\right)  \vspace{3mm}\\
\ds \phantom{G(\rho)\equiv}   +    \sum_{k=2}^N
(e^{r_k}-e^{-r_k})\left(
\frac{e^{\rho}-e^{-r_k}}{e^{-\rho}-e^{-r_k}}\prod_{l=2,\,l\ne k}^N
\frac{e^{-r_k}-e^{r_l}}{e^{-r_k}-e^{-r_l}}-1 \right) \ea \ee being
identically zero. Now $G(\rho)$ is obviously $2\pi i$-periodic,
and is readily seen to be entire. It is straightforward to check
that its limit for ${\rm Re}\,  \rho\to -\infty$ vanishes. Thus it
remains to verify that its limit for ${\rm Re}\,  \rho\to\infty$
is finite. Now the coefficient of $\exp (\rho)$ equals $F(\rho)-1$
(cf.~(\ref{Frho})), so it vanishes. Hence $G(\rho)$ vanishes for
${\rm Re}\,  \rho\to\infty$, too, so that (\ref{Csum}) is proved.

A second type of matrix playing an important role in this
paper is the Vandermonde matrix. Choosing
$\rho_1,\ldots,\rho_N\in{\mathbb C}$, it can be defined as the matrix
with $k$th row
\be\label{rowk}
\gamma_k\equiv \left(1,\rho_k,\ldots,\rho_k^{N-1}\right),\qquad
k=1,\ldots,N.
\ee
As is well known (and easily checked), it has determinant
\be\label{Vdet}
|{\rm Row}\,(\gamma_1,\ldots,\gamma_N)|=\prod_{1\le m<n\le
N}(\rho_n-\rho_m).
\ee
In the main text we encounter matrices of the form
\be\label{Vdm}
V(\rho_1,\ldots,\rho_N)\equiv {\rm
Row}\,(\eta_1,\ldots,\eta_N),\qquad \eta_k\equiv
\left(1-\rho_k,1-\rho^2_k,\ldots,1-\rho_k^N\right).
\ee
Presumably, the following result is known, too. In any case,
the proof we supply is short and simple.

\begin{lem}
One has \be\label{Vdet2} |V(\rho)|=\prod_{k=1}^N(1-\rho_k)\
\cdot\prod_{1\le m<n\le N} (\rho_n-\rho_m). \ee
\end{lem}

\noindent
{\bf Proof.} We may write
\be
\eta_k =(1-\rho_k)\left(1,1+\rho_k,\ldots, 1+\rho_k^2+\cdots
+\rho_k^{N-1}\right)\equiv (1-\rho_k)\zeta_k,
\ee
so that
\be
|V(\rho)|=\prod_{k=1}^N(1-\rho_k)\cdot |{\rm Row}\,
(\zeta_1,\ldots,\zeta_N)|.
\ee
In the matrix ${\rm Row}(\zeta_1,\ldots,\zeta_N)$ we now
subtract column $l$ from column $l+1$ for
$l=N-1,N-2,\ldots,1$, to obtain the Vandermonde matrix ${\rm
Row} (\gamma_1,\ldots,\gamma_N)$. Using (\ref{Vdet}), we
deduce (\ref{Vdet2}).\hfill \rule{3mm}{3mm}

\renewcommand{\thesection}{B}
\setcounter{equation}{0}
\setcounter{lem}{0}


\section*{Appendix B. Casorati determinants}

We begin by summarizing some known general results concerning
the solutions $F\in{\cal M}^{*}$ to the A$\Delta$Es at issue in this
paper, cf.~Ref.~\cite{norl}. Thus we have
\be\label{Fade}
F(x-i)+V_a(x)F(x+i)+(V_b(x)-c)F(x)=0,
\ee
where $c=\exp(p)+\exp(-p)$ is assumed to be a fixed
complex number for the moment. Assuming $F_1,F_2\in{\cal M}^{*}$
satisfy (\ref{Fade}), we define their Casorati determinant by
\be\label{casdef}
{\cal D}(F_1(x),F_2(x))\equiv F_1(x-i)F_2(x)-F_1(x)F_2(x-i).
\ee
Clearly, this function vanishes identically iff $F_1/F_2$
belongs to ${\cal P}_i$ (\ref{cPa}). Moreover, it satisfies the
A$\Delta$E
\be\label{casade}
{\cal D}(x)=V_a(x){\cal D}(x+i),
\ee
as is easily verified.


Assuming from now on $F_1/F_2\notin {\cal P}_i$, suppose
$F_3\in{\cal M}^{*}$ is a third solution to (\ref{Fade}). Then it
is straightforward to check that one has \be\label{F3}
F_3(x)=\nu_1(x)F_2(x)-\nu_2(x)F_1(x), \ee where\be\label{nuj}
\nu_j(x)\equiv {\cal D}(F_j(x),F_3(x))/{\cal
D}(F_1(x),F_2(x)),\qquad j=1,2. \ee Now it follows from
(\ref{casade}) that quotients of non-zero Casorati determinants
are $i$-periodic. Thus, one has $\nu_j\in{\cal P}_i$ whenever
$\nu_j\in{\cal M}^{*}$. Conversely, for $\nu_j\in{\cal P}_i$ the
rhs of (\ref{F3}) obviously solves (\ref{Fade}). Thus, the space
of meromorphic solutions to (\ref{Fade}) is two-dimensional over
the field of $i$-periodic meromorphic functions.

It should be repeated that the latter conclusion involves the
assumption that $F_1$, $F_2$ are solutions with a {\em non-zero}
Casorati determinant. To our knowledge, the {\em existence} of
such a basis is not known to follow from our assumptions
$V_a\in{\cal M}^{*}$, $V_b\in{\cal M}$, even when one also assumes the
asymptotics (\ref{Vas}). For the class of potentials obtained
above Theorem~2.3, however, we have solutions ${\cal W}(x,\pm p)$
available. (Here and from now on we take the $p$-dependence
into account.) We proceed by deriving more information
pertaining to the basis properties of the latter solutions.

\begin{theor}
Assuming
\be\label{pcondit}
e^p\ne e^{\pm r_1},\ldots,e^{\pm r_N},
\ee
we have
\be
{\cal D}({\cal W}(r,\mu;x,p),{\cal
W}(r,\mu;x,-p))=\left(e^p-e^{-p}\right)\lambda(r,\mu;x),
\ee
where $\lambda(r,\mu;x)$ is given by (\ref{laf}).
\end{theor}

\noindent
{\bf Proof}. It is clear from the definition (\ref{Vaf}) of
$V_a(x)$ and the A$\Delta$E (\ref{casade}) that
${\cal D}(x)/\lambda(x)$ is $i$-periodic. The problem is,
therefore, to show that this $i$-periodic function is the
constant $\exp(p)-\exp(-p)$. But we are going to prove this
by focusing on the $p$-dependence of the Casorati
determinant. Indeed, using (\ref{Wf}), it can be written
\be\label{casW}
\ba{l}
\ds
e^p\left( 1-\sum_{m=1}^N\frac{R_m(x-i)}{e^p-e^{-r_m}}\right)
\left( 1-\sum_{n=1}^N\frac{R_n(x)}{e^{-p}-e^{-r_n}}\right)
  \vspace{3mm}\\
\ds \qquad \qquad -e^{-p}\left(
1-\sum_{m=1}^N\frac{R_m(x)}{e^p-e^{-r_m}}\right)
\left( 1-\sum_{n=1}^N\frac{R_n(x-i)}{e^{-p}-e^{-r_n}}\right).
\ea
\ee
This shows that we are dealing with a $2i\pi$-periodic
meromorphic function of $p$, with simple poles in the period
strip ${\rm Im}\,  p\in (-\pi,\pi]$ that can be located only at $p=\pm
r_n$, $n=1,\ldots,N$, and with zeros for $p=0,i\pi$.

Consider now the residues at the pole $p=-r_k$. The first
factor in brackets has residue $-\exp(r_k)R_k(x-i)$ and the
third one $-\exp(r_k)R_k(x)$. The second factor in brackets
has value $d(r_k,\mu_k;x)R_k(x)$ (by virtue of the system
(\ref{sysN})), and, likewise, the fourth one has value
$d(r_k,\mu_k;x-i)R_k(x-i)$. Thus we obtain a total residue
\be
\ba{l}
\ds e^{-r_k}
\left(-e^{r_k}R_k(x-i)\right)d(r_k,\mu_k;x)R_k(x)
\vspace{2mm}\\
\ds \qquad \qquad -e^{r_k}\left(-e^{r_k}
R_k(x)\right)d(r_k,\mu_k;x-i)R_k(x-i)=0,
\ea
\ee
where we used (\ref{defd}).

We continue by observing that (\ref{casW}) is odd in $p$.
Hence the total residue for $p=r_k$ vanishes, too. We are now
in the position to conclude that the function obtained upon
dividing (\ref{casW}) by $\exp(p)-\exp(-p)$ is
$2i\pi$-periodic, entire and even in $p$. Its limit for ${\rm Re}\,
p\to \pm\infty$ equals
$1+\sum\limits_{n=1}^N\exp(r_n)R_n(x)=\lambda(x)$, so by Liouville's
theorem it equals $\lambda(x)$ for all $p$.
\hfill \rule{3mm}{3mm}

\medskip

From Theorem~B.1 we infer in particular that with the restriction
(\ref{pcondit}) in force, ${\cal W}(x,p)$ and ${\cal W}(x,-p)$
yield a basis in the above sense, as long as $\exp(p)$ is not
equal to $\pm 1$. It is indeed evident from (\ref{Wf}) that the
quotient \be {\cal W}(x,ik\pi )/{\cal W}(x,-ik\pi)=e^{-2k\pi
x},\qquad k\in{\mathbb Z}, \ee is $i$-periodic. However, for
$p=ik\pi$ one can obtain a further solution \be \lim_{\epsilon\to
0}(2\epsilon)^{-1}\left[{\cal W}(x,ik\pi +\epsilon)-e^{-2k\pi
x}{\cal W}(x,-ik\pi -\epsilon )\right]=\partial_p{\cal
W}(x,ik\pi),\quad k\in{\mathbb Z}.\hspace{-5mm}
 \ee Together with
${\cal W}(x,-ik\pi)$ this yields again a basis, since one clearly
gets \be {\cal D}(\partial_p{\cal W}(x,ik\pi),{\cal
W}(x,-ik\pi))=(-)^k\lambda(x). \ee

The auxiliary wave functions ${\cal A}(x,\pm p)$ (\ref{defaux}) are
well defined for all $p\in{\mathbb C}$, but of course their Casorati
determinant,
\be\label{casA}
{\cal D}({\cal A}(x,p),{\cal
A}(x,-p))=\left(e^p-e^{-p}\right)\lambda(x)\prod_{n=1}^N
\left( e^p-e^{-r_n}\right) \left( e^{-p}-e^{-r_n}\right),
\ee
vanishes when $\exp(p)$ equals $\exp(\pm r_m)$, cf.~also
(\ref{keyeq}) and (\ref{alm}). Just as for $\exp(p)=\pm 1$ one
can obtain solutions
\be
\lim_{\epsilon\to 0}(2\epsilon)^{-1}[{\cal A}(x,r_m+2ik\pi +\epsilon)
-\alpha_m(x){\cal A}(x,-r_m-2ik\pi -\epsilon
)]\equiv {\cal B}_{r_m,k}(x),
\ee
with $m=1,\ldots,N,k\in{\mathbb Z}$,
yielding a basis together with ${\cal A}(x,-r_m-2ik\pi)$. Indeed,
their Casorati determinants
\be
{\cal D}({\cal B}_{r_m,k}(x),{\cal A}(x,-r_m-2ik\pi))=
2^{-1}\partial_p{\cal D}({\cal A}(x,p),{\cal A}(x,-p))|_{p=r_m}
\ee
do not vanish identically.

\renewcommand{\thesection}{C}
\setcounter{equation}{0}
\setcounter{lem}{0}


\section*{Appendix C. Alternative representations}

Thus far, we have worked with formulas expressing $\lambda(x)$,
$V_a(x)$, $V_b(x)$ and ${\cal W}(x,p)$ in terms of the solution $R(x)$
to the system (\ref{sysN}), cf.~(\ref{laf}), (\ref{Vaf}),
(\ref{Vbf}) and (\ref{Wf}), resp. In this appendix we derive
alternative formulas involving various matrices and
determinants. They will be used in Appendix~D, as well as in
Parts~II and III.

We begin by deriving a second formula for the sum function
\be\label{sumf}
\Sigma(r,\mu;x)\equiv \sum_{n=1}^NR_n(r,\mu;x),
\ee
and hence for
\be
V_b(r,\mu;x)=\Sigma(r,\mu;x-i)-\Sigma(r,\mu;x),
\ee
cf.~(\ref{Vbf}).

\begin{theor}
The function (\ref{sumf}) can be rewritten as
\be\label{sum2}
\Sigma(r,\mu;x)=\sum_{m,n=1}^N\left([D(r,\mu;x)+C(r)]^{-1}\right)_{mn}.
\ee
\end{theor}

\noindent
{\bf Proof.} Denoting the canonical scalar product on ${\mathbb C}^N$
by $(\cdot,\cdot)$, we clearly have
\be
\Sigma(x)=(\zeta,R(x)).
\ee
Invoking (\ref{sysN}), this can be written as
\be
\Sigma(x)=\left(\zeta,[D(x)+C]^{-1}\zeta\right),
\ee
which amounts to (\ref{sum2}).\hfill \rule{3mm}{3mm}

\medskip

The formula (\ref{sum2}) is particularly useful in Appendix~D. We
now derive a third formula for $\Sigma(x)$ that is crucial in
Part~II.  To state the latter formula, we introduce the vectors
\be \omega_n(r)\equiv
\left(e^{r_n}-e^{-r_n}\right)(C(r)_{1n},\ldots,C(r)_{Nn})^t,\qquad
n=1,\ldots,N. \ee Now we define the matrix $\Omega_n(r,\mu;x)$ as
the matrix obtained from $D(r,\mu;x)+C(r)$ when the $n$th column
is replaced by $\omega_n(r)$.

\begin{theor}
The function (\ref{sumf}) can be rewritten
as
\be\label{sum3}
\Sigma(r,\mu;x)=\sum_{n=1}^N|\Omega_n(r,\mu;x)|/
|D(r,\mu;x)+C(r)|.
\ee
\end{theor}

\noindent
{\bf Proof.} In view of (\ref{sumf}) and (\ref{Rnform}), we
need only show
\be
\sum_{n=1}^N|Z_n(x)|=\sum_{n=1}^N|\Omega_n(x)|.
\ee
To this end we first replace the quantities $d(r_n,\mu_n;x)$
in the diagonals of the $2N$ matrices occurring here by
$\lambda_n\in{\mathbb C}$, and denote the resulting matrices by
$\hat{Z}_n(\lambda_1,\ldots,\lambda_N)$ and
$\hat{\Omega}_n(\lambda_1,\ldots,\lambda_N)$. Then it clearly
suffices to show
\be\label{ZOmh}
\sum_{n=1}^N|\hat{Z}_n(\lambda)|=\sum_{n=1}^N|\hat{\Omega}_n(
\lambda)|,
\ee
for arbitrary $\lambda\in{\mathbb C}^N$.

To this end we compare the coefficients of the monomials
$\lambda_{i_1}\cdots\lambda_{i_k}$ with $1\le i_1<\cdots
<i_k\le N$. For $k=0$ they are obtained by taking
$\lambda_1,\ldots,\lambda_N=0$. Expanding
$|\hat{Z}_n(0,\ldots,0)|$ with respect to its $n$th column
$\zeta$, we see that we get on the lhs the sum of all the
cofactors of the Cauchy matric $C(r_1,\ldots,r_N)$.
Similarly, expanding $|\hat{\Omega}_n(0,\ldots,0)|$ with
respect to its $n$th column $\omega_n$, we obtain on the rhs
$\sum\limits_{n=1}^N(\exp(r_n)-\exp(-r_n))|C(r_1,\ldots,r_N)|$. By
virtue of (\ref{Csum}), these two sums are equal.

Next, consider the case $k=N$. Since $\lambda_n$ does not
occur in $\hat{Z}_n(\lambda)$ and $\hat{\Omega}_n(\lambda)$,
this case gives rise to zero coefficients on the lhs and rhs.
Thus we are left with the case $k\in \{ 1,\ldots,N-1\}$.
Denoting the indices complementary to $i_1,\ldots,i_k$ by
$j_1,\ldots,j_l$, $l=N-k$, we should show that (\ref{ZOmh})
holds true when $\lambda_{j_1},\ldots,\lambda_{j_l}$ vanish.
We may restrict the summation to $n=j_1,\ldots,j_l$. Doing
so, we should choose the diagonal elements
$\lambda_{i_1},\ldots,\lambda_{i_k}$ in the expansion of the
determinants so as to get the pertinent coefficients. When
we then expand again with respect to the special columns, we
obtain on the lhs the sum of all cofactors of the
$(N-k)\times (N-k)$ matrix $C(r_{j_1},,\ldots,r_{j_l})$, and
on the rhs $|C(r_{j_1},\ldots,r_{j_l})|$ times the sum of
$\exp(r_n)-\exp(-r_n)$ with $n\in\{ j_1,\ldots,j_l\}$. Thus
the required coefficient equality follows once again from
(\ref{Csum}).\hfill \rule{3mm}{3mm}

\medskip

Consider next the function
\be\label{Q}
Q(r,\mu;x,p)\equiv
1-\sum_{k=1}^N\frac{R_k(r,\mu;x)}{e^p-e^{-r_k}}.
\ee
Evidently, it equals the wave function ${\cal W}(r,\mu;x,p)$
(\ref{Wf}) up to multiplication by the plane wave $\exp(ixp)$.
We proceed by obtaining a second
representation for
$Q$ as  a determinant quotient. As a preparation we introduce
the matrix
\be\label{defDe}
\Delta(r;p)\equiv {\rm diag}\, (\delta(r_1;p),\ldots,\delta(r_N;p)),
\ee
where
\be\label{defde}
\delta(\rho;p)\equiv 1-\frac{e^{\rho}-e^{-\rho}}{e^p-e^{-\rho}},
\ee
and the quotient function
\be\label{Qt}
\tilde{Q}(r,\mu;x,p)\equiv
|D(r,\mu;x)+C(r)\Delta(r;p)|/|D(r,\mu;x)+C(r)|.
\ee

For $N=1$ we have using (\ref{Cr})
\be
(D(x)+C)\tilde{Q}(x,p)=D(x)+C-C\frac{e^{r_1}-e^{-r_1}}
{e^p-e^{-r_1}}=D(x)+C-\frac{1}{e^p-e^{-r_1}},
\ee
whereas (\ref{Q}) and (\ref{Rnform}) yield
\be
(D(x)+C)Q(x,p)=D(x)+C-\frac{1}{e^p-e^{-r_1}}.
\ee
Therefore $Q(x,p)$ and $\tilde{Q}(x,p)$ are equal for $N=1$.

We are now going to prove by induction on $N$ that $Q$ and
$\tilde{Q}$ are equal for arbitrary $(r,\mu)$. For brevity we
use the notation
\be
F^{(M)}(\cdot)\equiv
F(r_1,\ldots,r_M,\mu_1,\ldots,\mu_M;\cdot),
\qquad M\in{\mathbb N} ,
\ee
in the proof. As a further preparation, we recall that
$Z_n(x)$ denotes the matrix obtained from $D(x)+C$ when the
$n$th column is replaced by $\zeta =(1,\ldots,1)^t$. The
identity
\be\label{Mid}
\left|Z^{(M)}_M(x)\right|=-\sum_{k=1}^{M-1}C^{(M)}_{Mk}\left|Z^{(M-1)}_k(x)\right|+
\left|D^{(M-1)}(x)+C^{(M-1)}\right|
\ee
is a cruci
al ingredient of our proof. It can be checked by
developing the determinant on the lhs w.r.t.~the last row.

\begin{theor}
One has
\be
Q(r,\mu;x,p)=\tilde{Q}(r,\mu;x,p),
\ee
for all $(r,\mu)$ satisfying (\ref{req1}), (\ref{req2}) and
(\ref{req3}).
\end{theor}

\noindent {\bf Proof}. We have already shown equality for $N=1$.
Assuming equality for $N=1,\ldots$, $M-1$, we now prove equality
for $N=M$.

We begin by noting that the system (\ref{sysN}) entails
\be\label{Qn}
Q^{(M)}(x,r_n)=d(r_n,\mu_n;x)R^{(M)}_n(x),\qquad n=1,\ldots,M.
\ee
Now we assert that we have
\be\label{QtQ}
\tilde{Q}^{(M)}(x,r_M)=Q^{(M)}(x,r_M).
\ee
To substantiate this assertion, we introduce
\be\label{G1}
G(x)\equiv
d(r_M,\mu_M;x)^{-1}\left|D^{(M)}(x)+C^{(M)}\right|\tilde{Q}^{(M)}(x,r_M),
\ee
and use (\ref{Qt}) to get
\be
G(x)=d(r_M,\mu_M;x)^{-1}\left|D^{(M)}(x)+C^{(M)}\Delta^{(M)}(r_M)\right|.
\ee
We now note the key property $\delta(\rho;\rho)=0$,
cf.~(\ref{defde}). It implies that the $M$th column of the
matrix on the rhs reads $(0,\ldots,0,d(r_M,\mu_M;x))^t$.
Developing its determinant w.r.t.~the $M$th column, we then
obtain
\be\label{step}
\ba{l}
\ds G(x)    =   \left |D^{(M-1)}(x)+C^{(M-1)}\Delta^{(M-1)}(r_M)\right|
\vspace{3mm}\\
\ds \phantom{G(x)}
 =    \tilde{Q}^{(M-1)}(x,r_M)\left|D^{(M-1)}(x)+C^{(M-1)}\right|.
\ea
\ee

We are now in the position to use the induction assumption.
It entails that (\ref{step}) can be rewritten as
\be
\ba{l}
\ds G(x)   =    Q^{(M-1)}(x,r_M)\left|D^{(M-1)}(x)+C^{(M-1)}\right|
\vspace{3mm}\\
\ds   \phantom{G(x)}=   \left|D^{(M-1)}(x)+C^{(M-1)}\right|-
\sum_{k=1}^{M-1}\frac{\left|Z_k^{(M-1)}(x)\right|}{e^{r_M}-e^{-r_k}},
\ea
\ee
where we used (\ref{Q}) and (\ref{Rnform}) in the second step.
Invoking the identity (\ref{Mid}), and using then (\ref{Qn}) and
(\ref{Rnform}), we infer
\be\label{G2}
G(x)=\left|Z^{(M)}_M(x)\right|=
d(r_M,\mu_M;x)^{-1}\left|D^{(M)}(x)+C^{(M)}\right|Q^{(M)}(x,r_M) .
\ee
Comparing (\ref{G1}) and (\ref{G2}), we obtain the asserted
equality (\ref{QtQ}).

More generally, it now follows that we have
\be\label{QQn}
\tilde{Q}^{(M)}(x,r_n)=Q^{(M)}(x,r_n),\qquad n=1,\ldots,M.
\ee
Indeed, this can be reduced to the case $n=M$ (already proved)
by relabeling $r_n,\mu_n$.

We continue by noting that (\ref{defDe})--(\ref{Qt}) entail
$\tilde{Q}^{(M)}(x,p)$ is of the form
\be
\sum_{i_1,\ldots,i_M=0}^1c_{i_1\cdots
i_M}(x)\prod_{k=1}^M \left(e^p-e^{-r_k}\right)^{-i_k},
\ee
with $c_{0\cdots 0}(x)=1$. Thus the function
\be
P(x,p)\equiv \prod_{k=1}^M \left(e^p-e^{-r_k}\right)\cdot
\left(Q^{(M)}(x,p)-\tilde{Q}^{(M)}(x,p)\right)
\ee
is a polynomial in $z=\exp(p)$ of degree at most $M-1$. Since
$P(x,p)$ vanishes in $M$ distinct points
$z=\exp(r_1),\ldots,\exp(r_M)$ (due to (\ref{QQn})), it
must be zero, entailing
$Q^{(M)}(x,p)=\tilde{Q}^{(M)}(x,p)$.
\hfill \rule{3mm}{3mm}

\medskip

Finally, let us observe that (\ref{Q}) and (\ref{laf}) entail
\be\label{Qla}
\lim_{{\rm Re}\,  p\to -\infty}Q(r,\mu;x,p)=\lambda(r,\mu;x).
\ee
We can therefore use Theorem~C.3 to obtain a second
representation for $\lambda(x)$, and hence for
$V_a(x)$ as well.

\begin{theor}
The functions (\ref{laf}) and (\ref{Vaf}) can be rewritten as
\be\label{la2}
\lambda(r,\mu;x)=\prod_{k=1}^N e^{2r_k}\cdot
|D(r,\mu;x-i)+C(r)|/|D(r,\mu;x)+C(r)|,
\ee
\be\label{Va2}
V_a(r,\mu;x)=|D(r,\mu;x-i)+C(r)||D(r,\mu;x+i)+C(r)|/
|D(r,\mu;x)+C(r)|^2.
\ee
\end{theor}

\noindent {\bf Proof.} From (\ref{defde}) we obtain \be \lim_{{\rm
Re}\,  p\to -\infty}\delta(\rho;p)=e^{2\rho}, \ee so that
(\ref{Qt}) yields \be \ba{l} \ds \lim_{{\rm Re}\,  p\to
-\infty}\tilde{Q}(x,p)    = \left|D(x)+C\,{\rm
diag}\left(e^{2r_1},\ldots,e^{2r_N}\right)\right|/
|D(x)+C|  \vspace{2mm}\\
\ds \phantom{ \lim_{{\rm Re}\,  p\to -\infty}\tilde{Q}(x,p)} =
\prod_{k=1}^N e^{2r_k}\cdot \left|D(x)\,{\rm
diag}\left(e^{-2r_1},\ldots,e^{-2r_N}\right)+C\right|/|D(x)+C|.
\ea\hspace{-10mm} \ee Recalling the definition
(\ref{defD})--(\ref{defd}) of $D(x)$, we deduce that $D(x)\,{\rm
diag}\left(e^{-2r_1},\ldots,e^{-2r_N}\right)$ equals $D(x-i)$.
Using Theorem~C.3, we then obtain \be \lim_{{\rm Re}\,  p\to
-\infty}Q(x,p)=\prod_{k=1}^Ne^{2r_k}\cdot |D(x-i)+C|/|D(x)+C|. \ee
Comparing this to (\ref{Qla}), one reads off (\ref{la2}). Now
(\ref{Va2}) is plain from (\ref{Vaf}). \hfill \rule{3mm}{3mm}

\renewcommand{\thesection}{D}
\setcounter{equation}{0}
\setcounter{lem}{0}


\section*{Appendix D. Formal self-adjointness}

Consider the complex translations $T_{\pm i}$ in the A$\Delta$O
$A$ (\ref{defA}). They have a quite natural interpretation as
self-adjoint operators on $L^2({\mathbb R},dx)$. Specifically, they can
be viewed as the Fourier transforms of multiplication by
$\exp(\pm p)$ on $L^2({\mathbb R},dp)$.

Next, letting $f\in{\cal M}$, we introduce the conjugate
meromorphic function
\be
f^{*}(x)\equiv \overline{f(\overline{x})}.
\ee
Now suppose that $V_a$ and $V_b$ satisfy
\be\label{sa1}
V_a^{*}(x)=V_a(x-i),
\ee
\be\label{sa2}
V_b^{*}(x)=V_b(x).
\ee
Then one readily checks that $A$ is formally
self-adjoint on
$L^2({\mathbb R},dx)$.

In this appendix we obtain conditions on $(r,\mu)$
entailing (\ref{sa1}) and (\ref{sa2}). Let us first note
that we have
\be
V_a^{*}(x)=\lambda^{*}(x)/\lambda^{*}(x-i),
\ee
\be
V_b^{*}(x)=\Sigma^{*}(x+i)-\Sigma^{*}(x).
\ee
Therefore, (\ref{sa1}) and (\ref{sa2}) will follow whenever
\be\label{last}
\lambda^{*}(x)=1/\lambda(x),
\ee
\be\label{sumst}
\Sigma^{*}(x)=-\Sigma(x-i),
\ee
resp.

To study (\ref{last}) and (\ref{sumst}), we exploit the
alternative formulas for $\Sigma(x)$ and $\lambda(x)$ obtained
in Appendix~C (cf.~(\ref{sum2}) and (\ref{la2})). We begin by
comparing
\be\label{s1}
\Sigma^{*}(x)=\sum_{m,n=1}^N\left(\left[D^{*}(x)+\overline{C}\right]^{-1}\right)
_{mn},
\ee
and
\be\label{s2}
-\Sigma(x-i)=\sum_{m,n=1}^N\left(\left[-D(x-i)-C\right]^{-1}\right)_{mn}.
\ee
Since $D(x)$ is a diagonal matrix, the formula (\ref{s1})
still holds true when we replace $\overline{C}$ by its
transpose $\overline{C}^t$. Now from (\ref{Cr}) we have
\be
\left(\overline{C}^t\right)_{kl}=\left[\exp(\overline{r}_l)-\exp(-
\overline{r}_k)\right]^{-1},
\ee
\be
(-C)_{kl}=\left[\exp(-r_l)-\exp(r_k)\right]^{-1}.
\ee
Thus we can ensure equality of $\overline{C}^t$ and $-C$ by
choosing $r_1,\ldots,r_N$ purely imaginary.

Doing so from now on, we conclude that (\ref{sumst}) will
follow when the multipliers are restricted in such a way that
we have
\be\label{Dst}
D^{*}(x)=-D(x-i).
\ee
From (\ref{defD}) and (\ref{defd}) we deduce that (\ref{Dst})
amounts to
\be\label{must}
\mu_n^{*}(x)=-\mu_n(x)e^{-2r_n},\qquad n=1,\ldots,N.
\ee
Since $r_n$ is purely imaginary, we need only introduce
multipliers $\hat{\mu}_n(x)$ by
\be
\hat{\mu}_n(x)\equiv ie^{-r_n}\mu_n(x),\qquad n=1,\ldots,N,
\ee
and require
\be\label{mut}
\hat{\mu}_n^{*}(x)=\hat{\mu}_n(x),\quad n=1,\ldots,N,
\ee
to obtain (\ref{must}).

The restrictions just obtained are also sufficient for
(\ref{last}) to hold true. Indeed, from (\ref{la2}) we have
\be
\lambda^{*}(x)=\prod_{n=1}^N \exp(2\overline{r}_n)\cdot
\left|D^{*}(x+i)+\overline{C}\right|/\left|D^{*}(x)+\overline{C}\right|.
\ee
Now we use the restriction $\overline{r}_n=-r_n$, (\ref{Dst})
and $\overline{C}=-C^t$ to obtain
\be
\lambda^{*}(x)    =    \prod_{n=1}^N
\exp(2\overline{r}_n)\cdot \left|-D(x)-C^t\right|/\left|-D(x-i)-C^t\right|
  =    1/\lambda(x).
\ee
(We used $|-M^t|=(-)^N|M|$ in the second step.)

We summarize the main result of this appendix in the following
theorem.

\begin{theor}
Assume that the numbers $r_n$, $n=1,\ldots,N$, are purely
imaginary, and that the functions $ie^{-r_n}\mu_n(x)$,
$n=1,\ldots,N$, are real-valued for real $x$. Then the potentials
$V_a(x)$ and $V_b(x)$ satisfy the formal self-adjointness
relations (\ref{sa1}) and (\ref{sa2}).
\end{theor}

\noindent
{\bf Proof.} This follows from the above reasoning. (Note
(\ref{mut}) is equivalent to real-valuedness of
$\hat{\mu}_n(x)$ for real $x$.)\hfill \rule{3mm}{3mm}

\medskip

It is an open question whether the above restictions
on $(r,\mu)$ entailing (\ref{sa1}) and (\ref{sa2}) are
necessary. In particular, our conditions imply that
$\lambda(x)$ is a phase factor for real
$x$ (cf.~(\ref{last})), a feature that is conceivably not
necessary for (\ref{sa1}) to be valid. At any rate, this
feature ensures that the formal operator
\be
V_a(x)T_{-i}=\lambda(x)T_{-i}\lambda(x)^{-1}
\ee
{\em may} be viewed as a unitary transform of the self-adjoint
operator $\exp(i\partial_x)$ on $L^2({\mathbb R},dx)$.

\subsection*{Acknowledgements}

Some of the work reported in this paper was done at the CIC
(Centro  Internacional de Ciencias) in Cuernavaca, Mexico. We
gratefully acknowledge the hospitality and financial support
provided by the CIC. We would also like to thank F~Calogero
for his invitation and illuminating discussions.

\subsection*{Note added in proof}

The uniqueness argument in the proof of Theorem~2.3 was inspired by a similar
argument in the context of reflectionless Schr\"odinger operators, which we
learned from Newell's monograph Ref.~\cite{newe}. Recently, E.~Date
informed us that he
has used the same type of arguments to handle soliton equations that admit a
Zakharov-Shabat formulation (in particular, the infinite Toda lattice), cf.
his papers \cite{date1} and \cite{date2}. This reasoning appears to have
been used for the
first time by Krichever in his theory of finite-gap solutions \cite{kric}.


\label{ruijsenaars-lastpage}


\begin{thebibliography}{99}
\small
\topsep0mm
\partopsep0mm
\parsep0mm
\itemsep0mm


\bibitem{r=0II} Ruijsenaars S N M, Reflectionless Analytic
Difference Operators II. Relations to Soliton Systems
(to appear in J. Nonlin. Math. Phys.).

\bibitem{r=0III} Ruijsenaars S N M, Reflectionless Analytic
Difference Operators III. Hilbert Space Aspects (to appear).

\bibitem{glf2} Ruijsenaars S N M,
Generalized Lam\'e Functions. II. Hyperbolic and Trigonometric
Specia\-li\-zations, {\it J. Math. Phys.}, 1999, V.40, 1595--1626.

\bibitem{hilb} Ruijsenaars S N M, Hilbert Space
Theory for Reflectionless Relativistic Potentials, to appear in
{\it Publ. RIMS Kyoto Univ.}.

\bibitem{NEEDS99} Ruijsenaars S N M,
Reflectionless Analytic Difference Operators
(A$\Delta$Os): Examples, Open Questions and Conjectures,  in
Proceedings NEEDS'99, {\it Suppl. to J. Nonlin. Math. Phys.}, 2001, V.8.

\bibitem{newe} Newell A C, Solitons in Mathematics and Physics,
SIAM, Philadelphia, 1985.

\bibitem{scot} Scott A C, Chu F Y F and McLaughlin D W,
The Soliton: a New Concept in Applied
Science, {\it Proc. IEEE}, 1973, V.61, 1443--1483.

\bibitem{toda} Toda M, Theory of Nonlinear
Lattices, Springer, Berlin, 1981.

\bibitem{cade} Calogero F and Degasperis A,
Spectral Transform and Solitons. Vol. I, North-Holland,
Amsterdam, 1982.

\bibitem{fata} Faddeev L D and Takhtajan L A,
Hamiltonian Methods in the Theory of Solitons,
Springer, Berlin, 1987.

\bibitem{aa3} Ruijsenaars S N M, Action-Angle Maps and
Scattering Theory for Some Finite-Dimensional Integrable
Systems III. Sutherland Type Systems and Their Duals,
{\it Publ. RIMS Kyoto Univ.}, 1995, V.31, 247--353.

\bibitem{norl} N\"orlund N E,  Vorlesungen \"uber
Differenzenrechnung, Springer, Berlin, 1924.

\bibitem{date1} Date E, On a direct method of constructing multi-soliton
solutions,
{\it Proc. Japan Ac.}, 1979, V.55, 27--30.

\bibitem{date2} Date E, Multi-soliton solutions and quasi-periodic solutions of
nonlinear equations of sine-Gordon type, {\it Osaka J. Math.}, 1982, V.19,
125--158.

\bibitem{kric} Krichever I M, Integration of nonlinear equations by the
methods of
algebraic geometry, {\it Func. Anal. Appl.}, 1977, V.11, 12--26.



\end{thebibliography}
\end{document}